\documentclass[11pt]{article}
\usepackage{graphicx}
\usepackage{pdfpages}
\usepackage{adjustbox}
\usepackage{amsmath,amsfonts,amssymb,amsthm}
\usepackage{caption}
\usepackage[ruled,vlined]{algorithm2e}
\usepackage{bbm}
\usepackage{breakcites}
\usepackage{tabularx}
\usepackage{geometry}
\usepackage{enumitem}
\usepackage{natbib}
\usepackage{bibunits} 
\usepackage{authblk}
\usepackage{graphicx}
\usepackage{subcaption}
\usepackage{hyperref}
\usepackage{ulem}
\usepackage{mathtools}

\defaultbibliographystyle{pa}
\defaultbibliography{bibtex_file,my,imai,sequential}

\setlist[itemize]{align=parleft,left=0pt..1em}
\newcolumntype{L}{>{\raggedright\arraybackslash}X}

\usepackage{stix}
\usepackage{color}

\DeclareMathOperator*{\argmin}{argmin} 

\newcommand{\dwh}[1]{{\color{black} {#1}}}

\theoremstyle{definition}
\newtheorem{definition}{Definition}[section]
\newtheorem{assumption}{Assumption}[section]
\newtheorem{example}{Example}[section]

\newtheorem{theorem}{Theorem}[section]

\newtheorem{lemma}[theorem]{Lemma}

\providecommand{\keywords}[1]
{
  \small	
  \textbf{{Keywords:}} #1
}

\makeatletter
\renewcommand\AB@affilsepx{, \protect\Affilfont}
\makeatother

    \title{Design-Based Confidence Sequences: A General Approach to Risk Mitigation in Panel Experiments}
    
    \author[1]{Dae Woong Ham} 
    \author[2]{Michael Lindon}
    \author[2]{Martin Tingley\footnote{Currently at Microsoft. This research was conducted while employed at Netflix.}}
    \author[3]{Iavor Bojinov}

    \affil[1]{\textit{Department of Statistics, Harvard University}}
    \affil[2]{\textit{Netflix, U.S.A}}
    \affil[3]{\textit{Harvard School of Business}}
  
    \date{   \vspace{-0.5cm} \today}
    
\begin{document}
\begin{bibunit}
\linespread{1.5}
    \maketitle
\vspace{-1cm}
\begin{abstract}
Randomized experiments have become the standard method for companies to evaluate the performance of new products or services. Beyond aiding managerial decision-making, experiments mitigate risk by limiting the proportion of customers exposed to innovations. Since many experiments are conducted sequentially over time, an emerging strategy to further derisk the process is to allow managers to ``peek'' at the results as new data become available and stop the test if the results are statistically significant. The class of statistical methods that allow managers to peek and still provide valid inference are often called anytime-valid since they maintain proper uniform type-1 error guarantees. In this paper, we extend existing anytime-valid approaches to accommodate the more complex yet standard settings in time series, switchback, and panel experiments. To achieve this, we leverage the design-based approach to focus on assumption-light and managerial relevant finite-sample estimands defined on the study participants as a direct measure of the risks incurred by companies. As a special case, our (asymptotic) results also provide a robust method for achieving always-valid inference in A/B tests. We further provide a variance reduction technique incorporating modeling assumptions and covariates. Finally, we demonstrate the effectiveness of our proposed approach through a simulation study and three real-world applications from Netflix. Our results show that using our confidence sequence, harmful experiments could be stopped after only observing a handful of units; for instance, our method would have stopped a 30,000 person Netflix experiment after the first 100 people.

\end{abstract}


\newpage
\section{Introduction}
Over the past decade, firms have widely adopted controlled experiments as the principled way of evaluating the performance of a new product or service (the treatment) relative to the current offering (the control). Researchers have found encouraging evidence that companies adopting control experimentation and developing a culture of experimentation see a boost in product innovation, identifying and scaling promising ideas, and informing strategic decision-making \citep{MS_decision_making2, MS_cohen, MS_cohen2, MS_koning, microsoft_online, dean_exp, paat_exp, mao2021quantifying}. For example, \citet{MS_koning} suggests that companies leveraging experimentation could see an increase of 30-100\% in revenue and visitation traffic after a year of initial adoption.

In addition to augmenting decision-making, experiments decrease the risk associated with innovation by limiting the initial proportion customers with access to the change \citep{kohavi2020trustworthy, risk_mitigation, li2024balancing}. This is especially relevant for changes designed to improve operational metrics, such as page load time, bandwidth consumption, and CPU utilization, where the experiment is used as a quality control rather than learning user preferences. In such cases, companies often use phased releases or rollouts to balance the deployment speed with the need for risk management \citep{li2024balancing,michael_paper2}. As large organizations often have millions of members, this small percentage still represents hundreds of thousands of users, providing sufficient power to detect even minor (negative) effects. However, if a treatment performs exceptionally poorly, prolonged exposure could lead to customer dissatisfaction and substantial revenue loss, a problem exacerbated by technology companies like Amazon, Netflix, and Microsoft conducting thousands of experiments on their customers each year \citep{thomke2020experimentation}. From a manager's perspective, it is vital to stop such negative experiments before they cause too much harm to the people in the study \citep{risk_mitigation}. For example, in a recent experiment conducted at Netflix (discussed in detail in Section~\ref{subsection:prepaid}), the manager planned to terminate the experiment if the number of lost customers due to the new version exceeded a specific threshold, such as 10,000 out of 500,000 potential customers in the study. In order to minimize risk, it is necessary to stop harmful experiments \textit{quickly}.

A potential solution involves allowing managers to continuously monitor and perform hypothesis tests on experimental results as new data becomes available and stopping the test if the results are significant. \dwh{Although this practice (often known as ``peeking'') is well known to invalidate the guaranteed type-1 error control \citep{Ramesh_anytime}, there exist methods that account for the peeking effect by widening the confidence interval, resulting in valid type-1 error control. This type of inference is often called ``anytime-valid'' inference because it allows researchers to sequentially test the same hypothesis at ``anytime'' while maintaining proper uniform type-1 error statistical guarantee at all times. However, a significant limitation of the existing research is the lack of methods that work beyond simple A/B tests and can work on time series, switchback, and panel experiments. For instance, switchback experiments are routinely implemented by marketplace companies, like Doordash, Lyft, and Uber, to quantify the impact of changes within a specific geographical area, like a city \citep{switchback, uber, jia_switchback, glynn_switchback, wager_switchback, xiong_switchback}.  In common switchback experiments, one experimental unit (as opposed to $n$ different individuals) is sequentially randomized to receive treatment or control for some pre-defined time, after which the outcomes are measured; this process is then repeated for $T$ periods. Time series experiments are the generalization of switchback experiments that allow the treatment assignment to dynamically update based on the observed data \citep{timeseries}. More generally, panel experiments --- the generalization of time series experiments to multiple experimental units --- have also grown in popularity as a way of increasing the personalization and power of experimentation \citep{panel_iav, tu_panel}. Prior work on anytime-valid inference does not directly apply to these settings because of the inherent intra-unit dependence that arises from the multiple observations on the same unit \citep{time_uniform, michael_paper}. Indeed, the problem is exacerbated whenever a unit's outcome at time $t$ depends on prior treatments; these types of carryover effects are common and can be long-ranged.}

In this paper, we focus on extending the current literature by proposing anytime-valid results that are also valid for time series and panel data experiments. Our proposed results focus on various causal estimands defined for the sample and are non-parameterically valid despite any non-stationarity or outcome dependencies through the design-based approach. A design-based approach uses the random treatment assignment as the basis of inference, which can equally be thought of as conditioning on the full set of (potential) outcomes, and performing inference on the sample. 
This approach has a long history dating back to Fisher and Neyman but has seen a resurgence in popularity as it allows us to perform inference on the realized sample and handle complicated experimental settings such as interference and non-stationarity with minimal assumptions \citep{fisher:1935, neym:35,timeseries, peng_luke_design, peng_anqi_DB, fredrik_DB}. In our time series context, this approach naturally accommodates any time series dependency or carryover effects without additional assumptions while performing inference on the realized sample. Our estimand is managerially relevant for capturing risk while our results still retain all the benefits of the existing anytime-valid approaches. Moreover, our results are agnostic to the outcome distribution, working for continuous, count, and binary data. This is particularly important as most companies compute the results across thousands of business metrics that have very different distributions. Finally, our confidence sequences are practically easy to compute as they only rely on means and variances, which can be efficiently computed for large streaming and batched data.

\subsection{Outline}
\label{subsection:outline_intro}
In Section~\ref{subsection:design_based_intro}, we \dwh{provide context on peeking and} introduce the design-based framework for the relevant causal estimands and estimators \dwh{in the time series setting}. In Section~\ref{section:nonasymp_intro}, we first give a brief overview of how confidence sequences are constructed via martingales. Then, we propose an exact confidence sequence that is valid for general settings but depends on potentially unknown parameters of the data-generating process. Section~\ref{section:main_section} introduces an improved design-based asymptotic confidence sequence. 
Section~\ref{section:TS_panel} generalizes the results to \dwh{the panel data setting where $n$ time series are observed}. We also provide illustrative examples of using our design-based confidence sequence for each of the aforementioned setting. \dwh{To emphasize our novel contributions, we start by presenting our results in the more complex time series setting. However, we still show how the simple typical A/B testing setting is a special case of our time series setting in Section~\ref{subsection:n_iid_setting}.}
Section~\ref{subsection:proxy_outcome} proposes an important variance reduction technique  by incorporating prior information or covariates. Section~\ref{section:simulations} shows simulations that demonstrate the favorable properties of our main results and contrasts confidence sequences with confidence intervals to guide practitioners. In Section~\ref{section:application}, we present three real world applications of our methodology to experiments that were run at Netflix. Finally, Section~\ref{section:discussion} presents our concluding remarks.


\section{\dwh{Confidence sequences and time series causal inference}}
\label{subsection:design_based_intro}
\subsection{Confidence intervals and sequences}
\label{subsection:intro_peeking}
\begin{figure}[t]
\begin{center}
\includegraphics[width=9cm]{"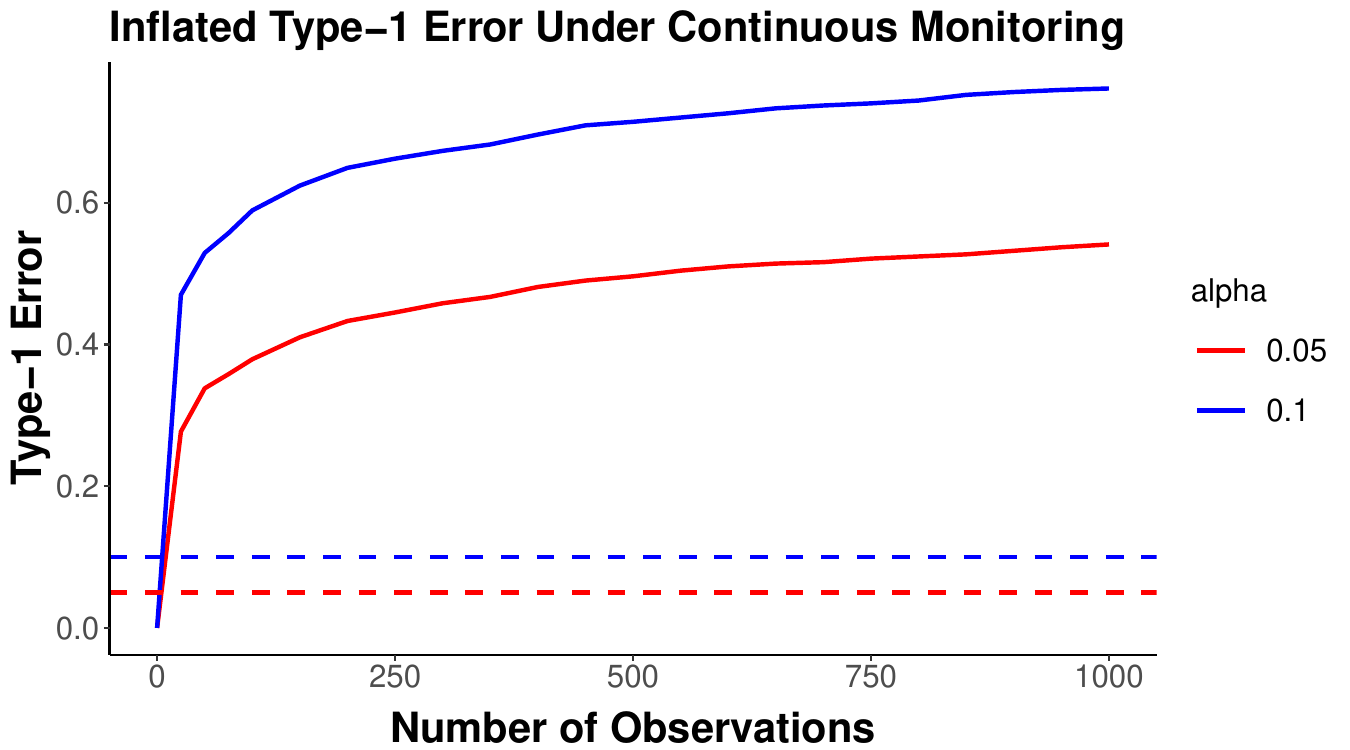"}
\caption{The figure plots the type-1 error when an analyst performs an $\alpha=0.05$ and $\alpha=0.1$ level $t$-test every time a new observation arrives ($x$-axis), when the data is independently and identically drawn from a $N(0,1)$ distribution. The solid red and blue lines show the inflated type-1 error rates as the number of observations grows for $\alpha = 0.05$ and $\alpha = 0.10$, respectively. The corresponding horizontal dotted lines show the expected type-1 error at $0.05$ and $0.10$, respectively.}
\label{fig:inflated_error}
\end{center}
\end{figure}

Unfortunately, performing multiple hypothesis tests (what is known informally as ``peeking'' at the data) invalidates the statistical guarantees when the experiment is designed for a single analysis at one fixed time. To formalize this, for a fixed $t$, $I_t$ is a valid confidence interval with type-1 error $\alpha$ for the true treatment effect $\mu_t$ iff 
\begin{equation}
\label{eq:CI}
\Pr( \mu_t \in I_t) \geq 1 - \alpha
\end{equation}
holds. Equation~\eqref{eq:CI} shows that the probability of $I_t$ containing $\mu_t$ is at least $1 - \alpha$, where $\alpha$ denotes the probability of failing to contain the truth (\emph{\emph{i.e.},} the type-1 error). Now, suppose we had a sequence of confidence intervals, $\{I_t\}_{t=1}^\infty$; Equation~\eqref{eq:CI} does not provide uniform type-1 error rate control. Specifically, for any $S \subset \mathcal{N}$, even if $I_t$ satisfies \eqref{eq:CI} for all $t\in S$, there is no reason to expect that $\Pr(\forall t \in S, \mu \in I_t) \geq 1- \alpha$. \dwh{Figure~\ref{fig:inflated_error} illustrates the inflated type-1 error when performing a $t$-test every time a new observation is observed, where each observation is independently drawn from a $N(0,1)$. For example, after $500$ observations, the type-1 error is uncontrolled at 70\% when $\alpha = 0.10$ as shown by the blue line. The issue is further exacerbated if the data come from a time series or panel experiment with adaptive assignment mechanism because of the additional correlation in the observations. }


To overcome the inflated type-1 error issue, researchers have proposed using methods to compute sequences of confidence intervals that allow analysts to continuously monitor the experiment \citep{michael_paper, nonasymp1, CS_ref1, CS_paper2, howard_nonasymp}. \dwh{For example, many companies, such as Adobe \citep{time_uniform}, Microsoft \citep{ian_bandit}, and Netflix \citep{michael_paper, michael_paper2}, have recently started incorporating the aforementioned methods in their experimentation platforms.} Confidence sequences are sequences of confidence sets with time-uniform coverage guarantees, that is, the probability that \textit{all} confidence sets cover the estimand is at least $1-\alpha$. Formally, a sequence of intervals $\{V_t\}_{t=1}^\infty$ is a $1-\alpha$ confidence sequence for a parameter $\mu_t$, \emph{\emph{e.g.}}, time-varying ATE, if
\begin{equation}
    \Pr(\forall t, \mu_t \in V_t) \geq 1 - \alpha \iff \Pr(\exists t, \mu_t \notin V_t) \leq \alpha.
\label{eq:validity}
\end{equation}
It follows that $\Pr(\mu_{T}\notin V_{T})\leq \alpha$ for arbitrary data-dependent stopping rule $T$. In words, Equation~\eqref{eq:validity} allows the analyst to continuously monitor ongoing experiments through $V_t$ and also immediately stop the experiment whenever the manager desires (for example detecting a significant negative treatment effect for the Netflix sign-up page experiment) while controlling the error below the nominal threshold, as depicted by the dotted blue and red lines in Figure~\ref{fig:inflated_error}, uniformly at \textit{all times}. 

\dwh{Additionally, anytime-valid inference enables companies to use any data-dependent stopping rules to terminate experiments early while still being statistically valid through Equation~\eqref{eq:validity}. This further allows} companies to rapidly and safely scale the number of experiments that are performed by automating the experiment lifecycle. For example, the manager can algorithmically terminate the experiment when the confidence sequence excludes zero, or when the confidence sequence is sufficiently small around zero indicating that any difference, if it exists, is not practically meaningful. As we show in our empirical application, this type of large-scale automation has the potential to significantly de-risk innovation.

\subsection{\dwh{Time Series Causal Estimands}}
In time series experiments, \dwh{\textit{one unit} is sequentially randomized into treatment or control at each time $t=1,\dots, T$. Let $W_t$ denote the treatment assignment at time $t$; we typically assume that $W_t$ is binary, with $W_t=0$ representing control and $W_t=1$ representing treatment. After the treatment assignment, we observe an outcome $Y_t$. Together, we observe two time series $\{W_t,Y_t\}_{t=1}^T$. In our empirical example, the unit of interest is a customer, the treatment is sending a message alert at day $t$ ($W_t = 1$) or not ($W_t = 0$). The outcome is then a measure of engagement $Y_t$, on day $t$. For all results in this paper, $T$ is allowed to be an arbitrary \textit{data-dependent} stopping time, \emph{e.g.}, stopping the experiment as soon as a negative or positive treatment effect is detected. }  
In most applied settings, there are often strong carryover effects such that the potential outcome is not only a function of its current treatment assignment but of the whole treatment assignment path \citep{timeseries, panel_iav}. For example, consider a ride-sharing platform that is evaluating a new surge pricing algorithm; so, at each time $t$, all drivers and passengers in a geographical area are assigned to either the new algorithm $(W_t=1)$ or the old algorithm $(W_t=0)$, thus the experimental unit is the entire ride-sharing algorithm as opposed to each driver and/or passenger. In this setting, high prices can temporarily dissuade passengers from ordering a car at time t, which can increase demand at time $t+1$ \citep{switchback, uber, dean_interference, panos_intereference}. Under the potential outcomes formulation of causal inference, we allow such \textit{carryover effects} by assuming that \dwh{at each time $t$, $\{Y_t(w_{1:t})\}$ (as opposed to $Y_t(w_t)$),} corresponding to what would happen if the unit is assigned \dwh{to a treatment assignment path $w_{1:t} = (w_1, \dots, w_t).$} 

We can connect the potential outcomes to the observed outcome by noting that
\begin{equation}
\label{eq:obs_outcome}
 \dwh{Y_t = \mathbf{1}(W_{1:t} = w_{1:t}^{obs}) Y_t(w_{1:t}^{obs}) },   
\end{equation}
\dwh{under the assumption of no non-compliance that is trivially satisfied in our experimental setup throughout the whole paper.} \dwh{For the remainder of the paper for any variable $O_t$, we denote $O_{1:t} = (O_1, \dots, O_t)$.}


Throughout, we treat the potential outcomes as fixed and leverages the randomization due to the experiment for inference. Therefore, Equation~\eqref{eq:obs_outcome} shows that $Y_t$ is only random through $W_t$ because the potential outcomes are fixed constants. This is typically referred to as a design-based approach and can equivalently be derived by conditioning on the full set of potential outcomes \citep{abadie2020sampling, rubin_design1, guillaume_design}. To make this formal, \dwh{we first denote the entire collection} of potential outcomes up to time $t$ as
$$Y_{1:t}(\bullet) = \{Y_1(\bullet), Y_2(\bullet), \dots, Y_t(\bullet) \}, $$
where $Y_t(\bullet) = \{Y_t(w_{1:t}): w_{1:t} \in \{0, 1\}^t \}$ denotes the entire possible collection of potential outcome at time $t$. The design-based approach crucially conditions on $Y_t(\bullet)$ for all $t$. Consequently, we denote $\mathcal{F}_{t}$ as the sigma-algebra that contains $Y_t(\bullet)$ and  all observed data up to time $t$, $\{W_j, Y_j\}_{j = 1}^t$, conditional on the potential outcomes. 

\dwh{Generally, we can define the treatment effect at time $t$ for any arbitrary treatment paths $w_{1:t}, w_{1:t}'$,
$$\tau_t(w_{1:t}, w_{1:t}') := Y_t(w_{1:t}) - Y_t(w_{1:t}'). $$
However, as the number of potential outcomes grows exponentially with $t$, most $\tau_t(w_{1:t}, w_{1:t}')$ cannot be estimated from the observed data without imposing additional assumptions \citep{timeseries}. To see the difficulty of the problem, consider only $T = 3$ total time steps. There are $2^3 = 8$ total assignment path combinations $Y_3(1, 1, 1), Y_3(0, 1, 1), \dots, Y_3(0, 0 ,0)$. We only observe one treatment path $w_{1:3}^{obs}$, thus observing only one of the eight combinations. If we make no modeling assumptions on the relationships between the different potential outcomes, then it is not possible to make progress and estimate $\tau_t(w_{1:t}, w_{1:t}')$ for \textit{any} $w_{1:t}, w_{1:t}'$ from one treatment assignment observation. This problem is exponentially exacerbated with larger $T$.}

\dwh{To make progress, \cite{timeseries} proposed focusing} on the contemporaneous causal effect (CTE) that is a function of the observed treatment path.
\begin{definition}[Contemporaneous Causal Effect \citep{timeseries}]
\label{def:causal_estimand}
    $$\tau_t(w_{1:(t-1)}^{obs}) := Y_t(w_{1:(t-1)}^{obs}, 1) - Y_t(w_{1:(t-1)}^{obs}, 0), \quad \bar{\tau}_t \coloneqq \frac{1}{t} \sum_{j = 1}^t \tau_j(w_{1:(j-1)}^{obs})$$
\end{definition}

The causal estimand $\tau_t(w_{1:(t-1)}^{obs})$\footnote{We specifically define the causal estimand as a function of our observed treatment path to show that our causal estimand changes as a function of our treatment path. Furthermore, defining the treatment effect in this way is similar to focusing on the average effect on the treated (ATT), which is a widely accepted causal estimand that changes as a function of the data \citep{rubin:imbens}.}, captures the short-term treatment effect of the unit receiving one additional treatment relative to control, conditional on the past. Typically, we focus on the temporal average of the CTE, \dwh{$\bar{\tau}_t$. Note that although companies often wish to estimate the total treatment effect had they deployed the treatment for all times versus control, \emph{i.e.}, $\tau_t^{\text{total}} := Y_t(1, 1, \dots, 1) - Y_t(0, 0, \dots, 0)$, this estimand is impossible to estimate without making strong (often parametric) assumptions about the outcome processes. Therefore, we focus on the CTE as a meaningful and attractive proxy for $\tau_t^{\text{total}}$ that can be non-parametrically estimated from the observed data \citep{timeseries}. Note that in the special case when there are no carryover effects, the CTE reduces to the familiar average treatment effect (ATE), $Y_t(1) - Y_t(0)$, and is exactly equivalent to $\tau_t^{\text{total}}$. }   

\dwh{Furthermore, there are several generalizations of the CTE discussed in \cite{timeseries} that can reduce the dependency on the observed assignment path by taking an expectation over it, also known as the ``stepping-approach''. Additionally, we can impose limits on the duration of the carryover effects by assuming that $Y_t(w_{1:t})$ only depends on the last $p$ treatment assignments, allowing us to write the potential outcomes as  $Y_t(w_{(t-p + 1):t})$. Section 3 of \citep{timeseries} details how to incorporate this assumption to estimate more general causal effects such as $\tau_t^{\text{total}}$. For brevity, we focus on the CTE without making any assumptions on the carryover effect. However, all our results can be generalized to the aforementioned generalization of the CTE, making our paper practically relevant for most meaningful estimands.} 

\subsection{\dwh{Estimation and Inference}}
Although the individual treatment effect at time $t$, $Y_t(w_{1:(t-1)}^{obs}, 1) - Y_i(w_{1:(t-1)}^{obs}, 0)$, is never jointly observed, we use the inverse propensity score estimator \citep{IPW, rubin:imbens} $\hat\tau_{t \mid t-1}$ that is unbiased for the individual treatment effect, where 
\begin{equation}
    \hat\tau_{t \mid t-1} \coloneqq \frac{\mathbbm{1}\{W_t = 1 \} Y_t}{p_{t \mid t-1}(1)} - \frac{\mathbbm{1}\{W_t = 0 \} Y_t}{p_{t \mid t-1}(0)}
    \label{eq:IPW}
\end{equation}
and \dwh{$p_{t \mid t-1}(w) \coloneqq \Pr(W_t = w \mid \mathcal{F}_{t-1})$. We purposefully allow the treatment assignment to be fully adaptive based on the past data $\mathcal{F}_{t-1}$, as is the case when multi-armed bandits are used. This allows our results to accommodate any adaptive time series experiments, where the propensity scores themselves are changing. For example, in our empirical application evaluating messaging, the company used an adaptive treatment allocation to learn how individuals responded to these message alerts in the past and updated the assignment probabilities of sending another message or not at day $t$ accordingly. Furthermore, this also allows us to tackle the common bandit settings, where an agent is typically tasked to maximize reward and pull the next arm as a function of all the previous treatment assignments \citep{MAB_overview, RLbook}.} 

To ensure that $\hat\tau_t$ is well defined, we make the following positivity assumption.
\begin{assumption}[Probabilistic Treatment Assignment]
\label{assumption:PTA}
There exists an $\epsilon > 0$ such that for every $t \in \mathbb{N}$, $w \in \{0, 1\}$ the following holding almost surely 
$$0 + \epsilon < p_{t \mid t-1}(w) < 1 - \epsilon$$
\end{assumption}
\noindent In words, Assumption~\ref{assumption:PTA} states that the propensity scores for every individual is bounded away from zero and one in the limit. We require the $\epsilon$ parameter since our main results hold asymptotically. Assumption~\ref{assumption:PTA} states that each unit arriving to the sign-up page has some positive probability to see both the new or current offering. Lastly, classical results show that the average of $\hat\tau_{t \mid t-1}$ is an unbiased estimator of $\tau_t(w_{1:(t-1)}^{obs})$, providing also a simple estimate of the (upper bound) of the variance. The variance estimate is an upper bound because the actual variance term contains the product $Y_i(w_{1:t}^{obs}, 1)Y_i(w_{1:t}^{obs}, 0)$, which are never jointly observed. Thus, we use the following inequality $a^2 + b^2 \geq 2ab$ for any $a,b \in \mathbb{R}$ to obtain the following upper bound. Throughout the rest of the paper we (with a slight abuse of wording) use ``variance estimate'' or ``estimated variance'' to refer to an estimate of the upper-bound of the variance (not the exact variance).
\begin{theorem}[Mean and Variance of Inverse Propensity Score Estimator]
\label{lemma:moment_cond}
Under Assumption~\ref{assumption:PTA},
$$\frac{1}{T}\sum_{t = 1}^T E(\hat\tau_{t \mid t-1} \mid \mathcal{F}_{t-1}) = \tau_t(w_{1:(t-1)}^{obs}).$$
Furthermore, the upper bound of the variance has the following unbiased estimator
\begin{equation}
\text{Var}(\hat\tau_{t \mid t-1} \mid \mathcal{F}_{t-1}) \leq E(\hat \sigma_{t \mid t-1}^2\mid \mathcal{F}_{t-1}), \quad  \hat\sigma_{t \mid t-1}^2 \coloneqq \frac{ \mathbbm{1}\{W_t = 1 \} Y_t^2}{p_{t \mid t-1}(1)^2} + \frac{\mathbbm{1}\{W_t = 0 \}Y_t^2}{p_{{t \mid t-1}}(0)^2},
\label{eq:sigma_estimate}
\end{equation}
where all expectations throughout the rest of this paper is taken with respect to the randomness of the treatment assignment conditioning on the potential outcomes.
\end{theorem}
The proof is provided in Appendix~\ref{appendix:proof_moments}.  Finally, we also assume the potential outcomes are bounded. 
\begin{assumption}[Bounded Potential Outcomes Under Carryover Effects]
$$ \limsup_{t} |Y_t(w_{1:t})| \leq M$$ for all $w_{1:t} \in \{0, 1\}^t$, $M \in \mathbb{R}$, where $\limsup$ is taken over all time points $t = 1, 2, \dots, \infty$.
\label{assumption:boundedPO_general}
\end{assumption}
\noindent The value of $M$ can be chosen to be extreme, ensuring that this assumption holds. This strategy is commonly employed to satisfy the regularity conditions necessary for design-based inference \citep{timeseries, peng_bounded}. In the finite-population framework, it is likely implausible for any units to have infinite potential outcome. Furthermore, Assumption~\ref{assumption:boundedPO_general} is about the \textit{realized} potential outcome. More concretely, this assumption states that the units allocated to the experiment have finite responses on the metrics used to evaluate the test.  Therefore, we view Assumption~\ref{assumption:boundedPO_general} as a mild regularity condition. Although the value of $M$ can be extreme, we show in Section~\ref{section:main_section} the value of $M$ has consequences on our exact confidence sequence but no impact on our main proposed asymptotic confidences sequence. 

\subsection{\dwh{Design-based approach}}
\label{subsection:DB_vs_SP}
\dwh{We now provide a brief overview of the design-based (also known as finite population) approach to analyzing randomized experiments and explain our rational for adopting this approach over the more common super-population approach to sequential testing \citep{Ramesh_anytime, time_uniform}. 

The fundamental difference between the finite-population and the super-population approach lies in the target estimand and the resulting inference assumptions \citep[Chapter~6]{rubin:imbens}. In the super-population approach, the experimental unit(s) are assumed to have been sampled (typically i.i.d.) from a hypothetical (infinite) super-population. The estimand of interest is then defined on the hypothetical super-population; for example, a standard estimand in this setting is the \textit{population average treatment effect}. In the finite population, there are no assumptions on the existence of a super-population, and the causal estimand is defined on the realized sample; for example, a standard estimand in this setting is the \textit{sample average treatment effect}. 

Under the design-based approach, the only source of randomness comes from the assignment mechanism, as the potential outcomes are considered to be fixed (unknown) quantities (equivalently, this can be thought of as conditioning on the full set of unobserved potential outcomes). Therefore, the observed outcome, $Y_t = \mathbf{1}(W_{1:t} = w_{1:t}^{obs}) Y_t(w_{1:t}^{obs})$, is random only through the randomness in the treatment assignment $W_{1:t}$, which is used for inference. Conversely, the super-population approach focuses on the randomness induced by the random sampling of unit(s), allowing us to model the potential outcome, $Y_t(w_{1:t}^{obs})$, as a random variable. The observed outcome is then random due to both the random treatment assignment and random potential outcome. The randomization is then only used to justify that treatment assignment is independent of all covariates rather than for inference.

Often, the super-population perspective is mathematically simpler, but the sampling assumptions can at times be more challenging to justify, whereas the finite-population approach makes no such assumptions but is often more mathematically complicated. Specifically, in the time series experiment setting, where a single unit is considered, there is no reasonable super-population (hypothetical or otherwise) that the inference can focus on, as was discussed in \cite{timeseries}. More formally, in order to achieve super-population inference, one must define what ``population'' the treatment effect  generalizes to, \emph{i.e.}, typically in a non-time series setting we usually make an i.i.d assumption and state that the potential outcome $(Y_i(1), Y_i(0)) \sim P$ (iid across $i$ individuals from a superpopulation of infinite units) for a common distribution $P$. In this setting, one can take the estimates learned from $n$ individuals and generalize that effect to a large super-population $P$ of infinite individuals. However, in a time series setting, where there are time serial correlations, we must specify the whole vector of potential outcomes $(Y_{1:T}(1), Y_{1:T}(0)) \sim \mathbf{P}_T$ follows a time series distribution $\mathbf{P}_T$. More importantly, we need to specify what $\mathbf{P}_T$ generalizes too, equivalent to what $P$ generalizes to. In this case there is no easy analogy to separate out the treatment effect from the time-series outcome process. Theoretically, we could try to generalize the ``time element'' $T$ to represent samples from a super-population process. For example, if $T$ was a 10 day process from Dec. 15 - Dec. 25, then we could generalize $\mathbf{P}_T$ to any 10 day process in the whole year, \emph{i.e.}, to Feb. 15 - Feb. 25. This is difficult to justify without strong parametric assumptions such as a weakly stationary process. To avoid this issue, most existing works on the design and analysis of time series experiments take the design-based approach \citep{panel_iav, timeseries, switchback} and focus on finite-sample estimands like the CTE in Definition~\ref{def:causal_estimand}.  

Moreover, since our paper is motivated to develop methods that can stop experiments when the risk of the study is deemed too large, what practically matters is the realized cost on the experimental units, not the cost on a hypothetical super-population. For example, managers often want to stop an experiment if the treatment is adversely impacting customers' engagement, as measured on the customer in the study by the CTE. Instead, the super-population estimand, $$\tau^{sp} := E[Y_t(w_{1:t-1}^\text{obs},1) - Y_t(w_{1:t-1}^\text{obs},0)]$$ takes an expectation with respect to the hypothetical distribution from which the unit was sampled, the interpretation is then that the manager would stop the experiment as soon the population level cost was deemed to be significant.  Practically, this is rarely the stopping criteria managers wish to apply.

Despite these differences, the point estimates of the treatment effect are the same as the inverse propensity score point estimator in Equation~\eqref{eq:IPW}. However, the variance expressions are not only different but also require different assumptions to be valid for their respective estimands. In particular, estimating $\text{Var}(Y_t(w_{1:t}))$ requires accounting for the additional source of randomness from the potential outcomes $Y_t(w_{1:t})$ in the super-population approach. In a typical randomized experiment, to justify using the sample variance estimator requires assuming that all treatment and control units have the same variance \citep{Ramesh_anytime}. In the time series setting, however, this is equivalent to assuming homoscedastic variance over time since there is only a single unit. This limiting assumption precludes the possibility of a non-stationary time series structure. If this assumption is false, the variance estimate would be invalid, resulting in invalid inference. Furthermore, to deal with outcome dependency, \citep{time_uniform} proposes a sample splitting approach to properly estimate the variances. On the other hand, the design-based approach bypasses all the aforementioned issues because $Y_t(w_{1:t})$ is constant, thus when estimating $\text{Var}(Y_t(w_{1:t}))$ we directly use $Y_t^2$ (as shown in the expression in $\hat\sigma_{t \mid t-1}^2$) for valid inference on the realized finite sample.

In summary, we chose the design-based approach to appropriately tackle the time series setting. Specifically, the design-based approach focuses on finite-population estimands appropriate for the one-unit time series experiment. Consequently, it also allows us to perform inference while tackling non-stationarity and carryover effects without making additional assumptions on the outcome $Y_t$ or time series structure. In exchange, the design-based approach does not make inference for an infinite/hypothetical population and formally makes valid inference for the obtained sample. However, we show that this in-sample estimand is also an estimand relevant for risk mitigation to managers. 
}

\section{Exact design-based confidence sequences}
\label{section:nonasymp_intro}

\subsection{Confidence sequence construction through martingales}
\label{subsection:intro_martingale}
We begin by introducing how confidence sequences are constructed through martingales. Equation~\eqref{eq:validity} requires that the confidence sequence $V_t$ contains the target parameter uniformly, allowing managers to perform hypothesis tests at \textit{every} time. On one hand, this can be viewed as a multiple-testing problem, where we control for type-1 error while performing multiple hypothesis tests. Although any naive approach such as Bonferroni correction will technically lead to valid type-1 error guarantees, such an approach leads to uncontrolled type-2 error since $\alpha \rightarrow 0$ as time grows. Thus, the state-of-the-art approaches have used martingale constructions that automatically give a time-uniform guarantee through Ville's maximal inequality \citep{vile}.
\begin{lemma}[Ville's Maximal Inequality]
\label{lemma:ville}
Let $Q_n$ be a non-negative supermartingale with respect to a filtration $\mathcal{F}$ with initial value $Q_0 = 1$. Then,
    $$\Pr\left(\exists n \in \mathbb{N}_0: Q_n \geq \frac{1}{\alpha}\right) \leq \alpha .$$
\end{lemma}
\noindent Contrasting Lemma~\ref{lemma:ville} with the desired guarantee in Equation~\eqref{eq:validity}, we see that Ville's Maximal Inequality directly gives the desired \textit{uniform} type-1 error guarantee. Thus, most constructions of confidence sequences rely on constructing a non-negative supermartingale and applying Lemma~\ref{lemma:ville} to achieve the time-uniform guarantee \citep{michael_paper, howard_nonasymp, nonasymp1}. In fact, recently \citep{martingale_must} show that all confidence sequences rely implicitly or explicitly on a martingale construction.

\subsection{Design-based exact confidence sequence}
\label{subsection:nonasympc_CS}
We now construct an exact (finite-sample valid) confidence sequence. Building upon existing works in \citep{ian_bandit, fan_lemma} and applying it to our time series setting, we utilize a truncated gamma mixture given by
$$f(\lambda) = \frac{\delta^\delta e^{-\delta \left(1 - \lambda\right)} \left(1 - \lambda\right)^{\delta - 1}}{\Gamma(\delta) - \Gamma(\delta, \delta)}$$
\dwh{for any $\delta >0$, where $\Gamma(., .)$ is the incomplete gamma function to help build our exact confidence sequence.}

\begin{theorem}[Design-based Non-asymptotic Confidence Sequence for the CTE with Gamma Mixture]
\label{theorem:nonasymp_CS_gamma_mixture}
Suppose $\{W_t, Y_t\}_{t = 1}^T$ are observed for arbitrary data dependent stopping time $T$, where Assumptions~\ref{assumption:PTA} and \ref{assumption:boundedPO_general} are satisfied. Let $m \coloneqq M/p_{min}$, where $p_{min} = \min\limits_{t, w} p_{t \mid t-1}(w)$, \emph{\emph{i.e.}}, $m$ is the most extreme value our estimate $\hat\tau_t$ can take for any $t$. Denote $S_{t \mid t-1} \coloneqq \sum_{j = 1}^t \hat\sigma_{j \mid j-1}^2$. Let the lower confidence sequence $L_t$ be defined as 
  $$L_t := \inf \{\tau_t(w_{1:(t-1)}^{obs}) : V_t(\tau_t(w_{1:(t-1)}^{obs}) \geq \frac{2}{\alpha} \},$$
where
$$V_t(\tau_t(w_{1:(t-1)}^{obs})) \coloneqq \left(\frac{\delta^\delta e^{-\delta}}{\Gamma(\delta) - \Gamma(\delta, \delta)}\right) \left(\frac{1}{B_t + \delta}\right) {_1F_1}(1, B_t + \delta + 1, A_t + B_t + \delta)$$ $$A_t \coloneqq \frac{\sum_{j = 1}^t \hat \tau_{t \mid t-1} - \tau_t(w_{1:(t-1)}^{obs}) }  {m}, \quad B_t \coloneqq \frac{S_{t \mid t-1}}{m^2}.$$
and ${_1F_1}(., .,.)$ is Kummer's confluent hypergeometric function. Similarly, define the upper confidence sequence as
$$U_t := 1 - \inf \{1 - \tau_t(w_{1:(t-1)}^{obs}) : V_t(1 - \tau_t(w_{1:(t-1)}^{obs}) \geq \frac{2}{\alpha} \}.$$
Then $(L_t, U_t)$ forms a valid $(1-\alpha)$ confidence sequence for the running mean of the contemporaneous average treatment effect $\bar{\tau}_t(w_{1:(t-1)}^{obs})$.
\end{theorem}
The proof is in Appendix~\ref{appendix:proof_extension} and follows previous proof techniques in \citep{ian_bandit}. There are two main practical limitations to Theorem~\ref{theorem:nonasymp_CS_gamma_mixture}. First, the confidence sequence scales with $M$, which is dependent on both the observed and missing potential outcomes. Although Assumption~\ref{assumption:boundedPO_general} can be seen as a mild regularity condition, analysts often do not know this extreme value, except in special cases such as binary outcomes. This issue is further exacerbated if there exists one extreme potential outcome. 

Second, in the spirit of a sequential test, the analyst may desire to change $p_{t \mid t-1}(w)$ as data arrives. However, Theorem~\ref{theorem:nonasymp_CS_gamma_mixture} requires $p_{min}$ to be determined before the start of the experiment, thus restricting the treatment assignment probabilities to a pre-specified range. Although this is common practice for many experiments, the confidence width further scales inversely with $p_{min}$. We remark that there exist more sophisticated methods that can bypass this issue with stronger concentration inequalities \citep{ian_bandit} that allow using the empirical variance as opposed to the minimum $p_{min}$. 

Despite these limitations, this confidence sequence is shown to have provably an asymptotic rate of $O(\sqrt{B_t \log(B_t)}/t)$, which is near optimal \citep{ian_bandit}. We present an alternative exact confidence sequence in Appendix~\ref{appendix:proof_nonasymp} that suffers from a suboptimal rate but has a closed-form expression. 


\section{Design-based asymptotic confidence sequences}
\label{section:main_section}
We now improve the design-based \textit{non-asymptotic} confidence sequence introduced in Section~\ref{subsection:nonasympc_CS} by relaxing assumptions and obtaining an \textit{asymptotically} valid confidence sequence. Asymptotic confidence sequences were first introduced by \citep{time_uniform}, and we extend their work to the design-based framework \dwh{by tackling more complex experimental settings in the time series and panel data settings}. 

\subsection{Asymptotic confidence sequence}
Informally, asymptotic confidence sequences are valid confidence sequences after a ``sufficiently large'' time. Although this may worry practitioners since confidence sequences, by definition, should have valid coverage at all times (including early times), we show through simulations in Section~\ref{section:simulations} that the empirical coverage remains robust in practice because the confidence sequence width for early times is wide in part due to our upper bound variance estimator.

\begin{definition}[Asymptotic Confidence Sequences \citep{time_uniform}]
\label{def:asymp_CS}
We say that ($\hat\mu_i \pm V_i$) is a two-sided $(1-\alpha)$ asymptotic confidence sequence for a target parameter $\mu_i$ if there exists a non-asymptotic confidence sequence $(\hat\mu_i \pm \tilde{V}_i)$ for $\mu_i$ such that 
\begin{equation}
\label{eq:asymp_CS_req}
\frac{\tilde{V}_t}{V_t} \xrightarrow{a.s.} 1.
\end{equation}
Furthermore, we say $V_t$ has an approximation rate $R$ if 
$\tilde{V}_t- V_t = O_{a.s.}(R)$, where $R$ can be interpreted as how fast the approximation error $\tilde{V}_t - V_t$ is decreasing.
\end{definition}
To readers familiar with asymptotic confidence \textit{intervals}, the above definition may be puzzling. To give some intuition, the idea of ``couplings'' has been used in the literature of strong approximations and invariance principals \citep{approx1, approx2} to formally define asymptotic confidence \textit{intervals}. This literature defines a $(1- \alpha)$ asymptotic confidence interval if there exists a \textit{non-asymptotic} (unknown) confidence interval centered at the same statistic such that the difference between this non-asymptotic and asymptotic confidence interval is negligible asymptotically. Equation~\eqref{eq:asymp_CS_req} captures the same notion except we replace all convergence statement with almost-sure convergence to satisfy the time \textit{uniform} guarantee required of confidence sequences.\footnote{This is further discussed and proven in Appendix C.4 of \citep{time_uniform}} Throughout the rest of the paper (including the appendix), we omit subscript ``a.s.'' from $o_{a.s.}(.)$ and $O_{a.s.}(.)$ to simplify notation.

\subsection{Design-based asymptotic confidence sequence for the CTE}
\label{subsection:first_result}
We now state one of the main result of our paper, namely that we can construct asymptotically valid confidence sequences for $\tau_t(w_{1:(t-1)}^{obs})$ using $\hat\tau_{t \mid t-1}$ and our estimated variance $\hat\sigma_{t \mid t-1}^2$. Before stating the theorem, we require one additional assumption so that our variance is well-behaved and does not vanish in the limit to allow for asymptotic approximations.
\begin{assumption}[Non-Vanishing Variance]
\label{assumption:var_no_vanish}
In Appendix~\ref{appendix:proof_moments}, we show that $\text{Var}(\hat\tau_{t \mid t-1}) \leq \sigma_{t \mid t-1}^2$, where 
\begin{equation}
\label{eq:sigma_exp}
    \sigma_{t \mid t-1}^2 =  \frac{Y_t(w_{1:(t-1)}^{obs}, 1)^2}{p_{t \mid t-1}(1)} + \frac{Y_t(w_{1:(t-1)}^{obs}, 0)^2}{p_{t \mid t-1}(0)}.
\end{equation}
We say Assumption~\ref{assumption:var_no_vanish} holds if
$$\frac{1}{\sum_{j = 1}^t \sigma_{t \mid t-1}^2} = o(1) \iff \tilde{S}_t \coloneqq \sum_{j = 1}^t \sigma_{j \mid j-1}^2 \xrightarrow{t \rightarrow \infty} \infty  \text{ almost surely.}$$
\end{assumption}
\noindent Assumption~\ref{assumption:var_no_vanish} holds if $\frac{1}{t} \tilde{S}_t \xrightarrow{a.s.} \sigma_{*}^2$ or if $\sigma_1^2 = \sigma_2^2 = \dots = \sigma_T^2$. Informally, Assumption~\ref{assumption:var_no_vanish} is satisfied as long as the potential outcomes do not vanish to zero as time grows. Conversely, one way Assumption~\ref{assumption:var_no_vanish} does not hold is if the response of \textit{all} the units arriving over time suddenly vanish to zero after a certain time. We believe this is unlikely in practice.

\begin{theorem}[Design-based Asymptotic Confidence Sequences for the Contemporaneous Treatment Effect with Carryover Effects]
Suppose $\{W_t, Y_t\}_{t = 1}^T$ are observed for arbitrary data dependent stopping time $T$, where Assumptions~\ref{assumption:PTA}-~\ref{assumption:var_no_vanish} are satisfied. Then,
$$ \frac{1}{t} \sum_{j = 1}^t \hat\tau_{j \mid j -1} \pm  \frac{1}{t} \sqrt{ \frac{S_{t \mid t-1} \eta^2 + 1}{\eta^2} \log \Bigg( \frac{S_{t \mid t-1} \eta^2 + 1}{\alpha^2}\Bigg) } $$
forms a valid $(1-\alpha)$ asymptotic confidence sequence for the running mean of the CTE, $\bar{\tau}_t(w_{1:(t-1)}^{obs})$, with approximation rate $o\left(\sqrt{\tilde{S}_{t \mid t-1} \log \tilde{S}_{t \mid t-1}}/t\right)$ for any pre-specified constant $\eta > 0$, where $S_{t \mid t-1}$ is defined in Theorem~\ref{theorem:nonasymp_CS_gamma_mixture}.
\label{theorem:main_general_thm}
\end{theorem}

The proof is in Appendix~\ref{appendix:proof_main_them} and follows from using strong martingale difference sequence approximation, constructing a martingale from the limiting Gaussian distribution, and plugging in a consistent variance estimator. In contrast to the non-asymptotic confidence sequence presented in Section~\ref{subsection:nonasympc_CS}, the confidence width has no expressions related to $p_{min}$ or $M$, thus making it a fully general result. Ignoring constant terms, Theorem~\ref{theorem:main_general_thm} also shows that the width scales similar to a confidence interval, which usually scales as $O(\sqrt{S_t}/t)$. However, the confidence sequence has an extra log penalty term to achieve the necessary time-uniform guarantees (see Section~\ref{subsection:sim2} for simulations comparing confidence intervals and sequences).

Lastly, $\eta$ is a tuning parameter typically chosen by the analyst to minimize the confidence width at a certain fixed time. Given $\tilde{S}_n$ and $t$, the confidence width can be minimized for a choice of $\eta$, and we give a closed form expression for the minimum in Appendix~\ref{Appendix:choosing_rho}. Although the choice of $\eta$ can impact the confidence sequence especially at the beginning when sample size is small, the choice of $\eta$ becomes less significant as more data arrives. Consequently, for all examples and simulations we choose $\eta$ so that the width is minimized at time 10 since the confidence sequence width is the largest at early times. We now illustrate Theorem~\ref{theorem:main_general_thm} through Example~\ref{example:carryover}.

\begin{figure}[t]
\begin{center}
\includegraphics[width=12cm]{"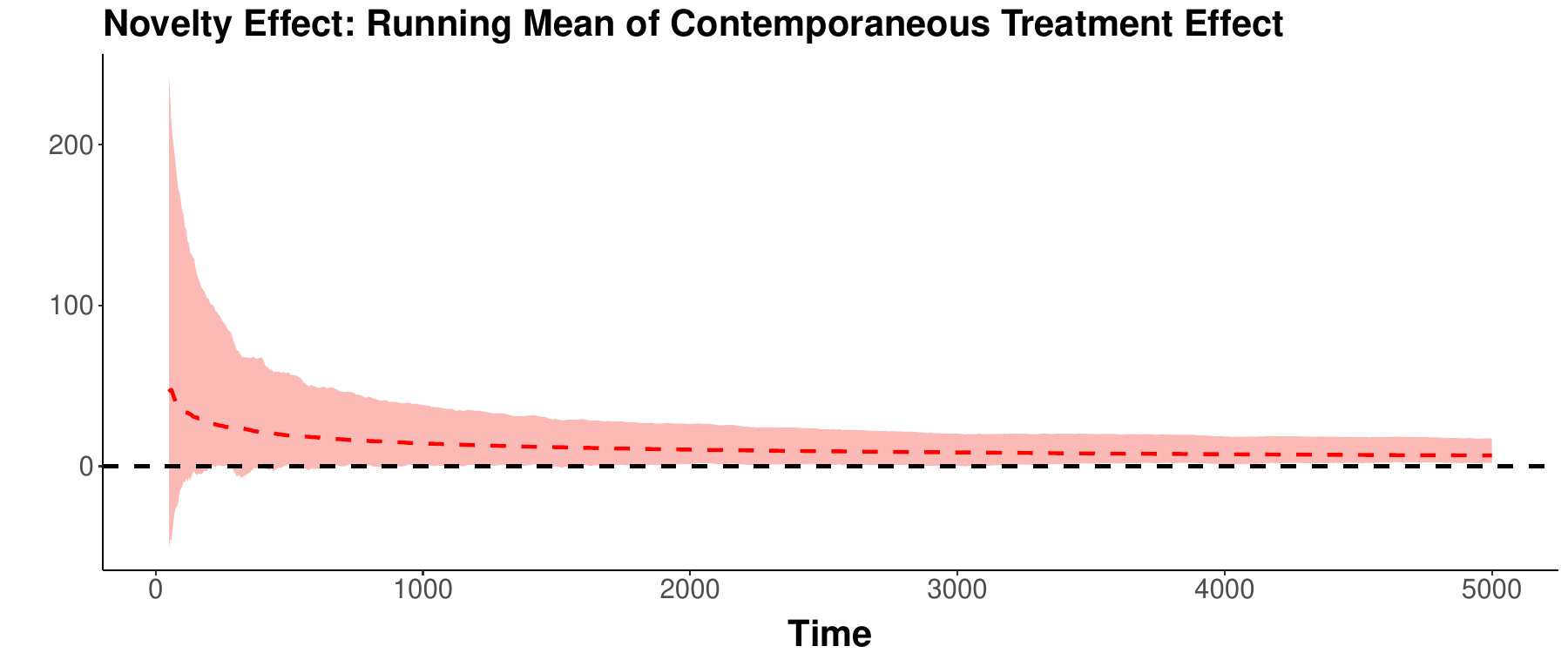"}
\caption{Novelty Effect of Message-Alert Treatment (Example~\ref{example:carryover}). The red contours show the lower and upper confidence sequence for the time varying novelty treatment effect of message alerts at $\alpha = 0.05$ using Theorem~\ref{theorem:main_general_thm}. The red dotted line represents the true time varying treatment effect that diminishes to zero and the black horizontal dotted line represents the zero (null) line. }
\label{fig:carryover_example}
\end{center}
\end{figure}

\begin{example}[Novelty Effect of Message Alerting Treatment]
\label{example:carryover}
Inspired by a Netflix case discussed later in Section~\ref{section:application}, suppose  one individual is experimented on across day $t$ and the treatment is whether to send a push notification over text or email to this individual at day $t$ ($W_t = 1$) or not ($W_t = 0$). We record whether the individual engaged with the product on day $t$ or not, \emph{i.e.}, $Y_t = 1$ if the individual engaged on day $t$. For simplicity, suppose the individual is randomly selected to receive a message at day $t$ or not with a fair $\text{Bern}(0.5)$ coin flip. Suppose the user's engagement increases substantially in the beginning based on the new treatment but the treatment effect diminishes over time (to zero) because the user has grown used to the treatment (often known as novelty effect). Furthermore, we assume that the message alerting treatment is only effective if the user did not receive a message alert at the previous time. Formally, we simulate data under the following model: 
$$Y_t(w_{t-1} = 0, w_t = 1) = Y_t(w_{t-1} = 0, w_t = 0) + 500/\sqrt{t}$$
and $Y_t(w_{t-1} = 0, w_t = 0) = Y_t(w_{t-1} = 1, w_t = 0) = Y_t(w_{t-1} = 1, w_t = 1) \sim N(25, 10^2)$, \emph{\emph{i.e.}}, there is indeed carryover effects. We exaggerate the initial treatment effect to start at 500, which decreases with order $1/\sqrt{t}$, to clearly show the time varying effect. Figure~\ref{fig:carryover_example} shows that the confidence sequence uniformly covers the true time varying running mean of the contemporaneous treatment effect (in red) at all times. Although the analyst would reject the null effect at approximately $t = 200$, the analyst would also see a diminishing time varying treatment effect had the analyst suspected a novelty effect and continued running the experiment.
\end{example}

\subsection{Average Treatment Effect for $N$ Non-Interfering Units}
\label{subsection:n_iid_setting}
\dwh{Before generalizing our results to the panel data setting, we briefly show how all our results carry through in the simpler case when estimating the ATE of $N$ units instead of one time series. Concretely,} suppose now we have $n$ units arriving sequentially. For each unit, we randomly assign each unit to a treatment $W_i$ and subsequently observe a corresponding outcome $Y_i$ so that we observe data $\{W_i, Y_i\}_{i = 1}^N$. \dwh{We remark that all notation is similar to the time series setting except we replace subscript time $t$ with individual units $i = 1, 2, \dots, n$.} 

Our goal is to estimate the finite-sample average treatment effect $\tau_N$ for the $N$ individuals in our obtained experimental sample.
\begin{definition}[Finite-Sample Average Treatment Effect]
\label{def:ATE}
    $$\tau_N := \frac{1}{N} \sum_{i = 1}^N Y_i(1) - Y_i(0),$$
\end{definition}
\noindent where $N$ denotes our finite-sample population of interest. In a sequential experimental setting, $N$ changes as more data arrives (further formalized in Theorem~\ref{theorem:nonasymp_CS_gamma_mixture}). In Definition~\ref{def:ATE}, we impose the no-interference assumption that is typically assumed in this setting as the $N$ individuals are independently arriving and do not affect each other's outcome. 

We keep all the same estimators, $\hat \tau_{t \mid t-1}, \hat \sigma_{t \mid t-1}^2$, but we replace subscript $t$ with $i$, \emph{\emph{i.e.}}, $\hat\tau_{i \mid i-1}, \hat \sigma_{i \mid i-1}^2$, respectively, to denote the estimate of the treatment effect and variance of each \textit{individual} $i$. \dwh{Because this setting is a simpler setting than the time series with carryover effects and dependencies across time, our results directly hold in this simpler setting when estimating the ATE in Definition~\ref{def:ATE}. For completeness, we formally present the analogue of our main Theorem~\ref{theorem:main_general_thm} as a Lemma in this setting. However, all results including results in later sections carry through similarly. }

\begin{lemma}[Design-based Asymptotic Confidence Sequence for the ATE]
Assume the same setting as that in Theorem~\ref{theorem:main_general_thm}\footnote{In Assumption~\ref{assumption:var_no_vanish} the expression of $\sigma_{t \mid t-1}^2$ contained $Y_t(w_{1:(t-1)}^{obs}, 1)^2, Y_t(w_{1:(t-1)}^{obs}, 0)^2$. For this setting, we replace those potential outcomes with $Y_i(1)^2, Y_i(0)^2$ since there is no carryover effect.} is satisfied. Then,
$$ \frac{1}{n} \sum_{i = 1}^n \hat\tau_i \pm  \frac{1}{n} \sqrt{ \frac{S_n \eta^2 + 1}{\eta^2} \log \Bigg( \frac{S_n \eta^2 + 1}{\alpha^2}\Bigg) }$$
forms a valid $(1-\alpha)$ asymptotic confidence sequence for the average treatment effect $\tau_n$ with approximation rate $o\left(\sqrt{\tilde{S}_n \log \tilde{S}_n}/n \right)$ for any pre-specified constant $\eta > 0$.
\label{theom:main_thm1_ATE}
\end{lemma}

\noindent We provide an example of applying Lemma~\ref{theom:main_thm1_ATE} to the multi-arm bandit setting in Appendix~\ref{appendix:n_iid_setting}.

\subsection{Panel data setting}
\label{section:TS_panel}
\subsubsection{Setting and notation}
The above theorems and results focus on the context of time series experiment. In practice, many organizations run multiple time series experiments for all $n$ fixed units, where we observe multiple responses $Y_{i,t}$ for each unit $i = 1, 2, \dots, n$ \citep{panel_iav}, combining the familiar ATE of $n$ units in Section~\ref{subsection:n_iid_setting} across time $t$.
In this setting, we also observe the treatment assignment $W_{i, t}$ for each unit $i$ across time $t$, where as before $w_{i, t}$ is a realization of this random variable and $w_{i, 1:t}$ denotes the entire treatment path for unit $i$ until time $t$. Although our above setting allows for any general carryover effects, in this setting we assume that each unit's potential outcome is only a function of its own treatment assignment path. More formally,
\begin{assumption}[Independence of Potential Outcome Across Units]
$Y_{i, t}(w_{-i, 1:t}, w_{i, 1:t}) = Y_{i, t}(w_{-i, 1:t}', w_{i, 1:t})$ for all $w_{-i, 1:t}, w_{-i, 1:t}'$,  $i = 1, 2, \dots, n$, and $t = 1, 2, \dots, T$, where $w_{-i, 1:t}$ denotes the treatment assignment paths for all units up to time $t$ except unit $i$.
\label{assumption:ind_PO}
\end{assumption}
Concretely, Assumption~\ref{assumption:ind_PO} states that one unit's treatment does not impact another, likely disconnected, unit's outcome. This would be violated, for example, if units in the experiment communicate about their Netflix experience to each other. This assumption allows us to denote each unit's potential outcome as $Y_{i, t}(w_{i, 1:t})$, where we again assume the potential outcome is also not a function of the future treatment assignments. As before, we denote the entire collection of potential outcome for all units at time $t$ as
$$Y_{n,t}(\bullet) = \{Y_{1,t}(w_{1, 1:t}), Y_{2,t}(w_{2, 1:t}), \dots, Y_{n,t}(w_{n, 1:t}): w_{i, 1:t} \in \{0, 1\}^t \}$$
and consequently define the entire collection of potential outcome for all units up to time $t$ as $Y_{n, 1:t}(\bullet) = \{Y_{n, 1}(\bullet), Y_{n, 2}(\bullet), \dots, Y_{n, t}(\bullet) \}$.

The new filtration $\mathcal{F}_{n, t}$ denotes the sigma algebra containing the information set of all treatment assignment $W_{i, 1:t}$ and observed outcome $Y_{i, t}$ up to time $t$ for all units $i$. Furthermore, the filtration always contains the entire potential outcome set $Y_{n, 1:T}(\bullet)$. Similar to Assumptions~\ref{assumption:PTA} and~\ref{assumption:boundedPO_general}, we assume all the potential outcomes are bounded and the treatment assignments are bounded away from zero or one, which we now state under one assumption.
\begin{assumption}[Bounded Potential Outcomes and Treatment Assignment in Panel Data Setting]
\label{assumption:panel_data_regularity}
$$\limsup_{i, t} Y_{i, t}(w_{i, 1:t}) \leq  M, \quad 0 + \epsilon < p_{i,t \mid t-1}(w) \coloneqq \Pr(W_{i,t} = w \mid \mathcal{F}_{n, t-1}) < 1 - \epsilon,$$
for $M \in \mathbb{R}$ and all $i = 1, 2, \dots, n$, $t = 1, 2, \dots, T$, and $w_{i, 1:t} \in \{0, 1\}^t$, where $\limsup$ is taken over all units $i$ and time periods $t$.
\end{assumption}
Assumption~\ref{assumption:panel_data_regularity} allows the experimenter to adapt the treatment assignments for unit $i$ not only as a function of the previous unit $i$'s treatment assignment and response but also all the other units' treatment assignment and response. Lastly, since we have multiple units $n$, our causal estimand changes to the contemporaneous treatment effect averaged over $n$ units:
\begin{equation}
\label{eq:ATE}
\tau_{n,t}(w_{1:n, 1:(t-1)}^{obs}) := \frac{1}{n} \sum_{i = 1}^{n}Y_{i, t}(w_{i, 1:(t-1)}^{obs}, 1) - Y_{i,t}(w_{i, 1:(t-1)}^{obs}, 0),
\end{equation}
where $w_{1:n, 1:(t-1)}^{obs}$ denotes the entire observed treatment path for all units $i = 1, 2, \dots, n$ up to time $t- 1$. As a reminder, all causal estimands in Equation~\eqref{eq:ATE} and Definition~\ref{def:causal_estimand} also capture quantities directly related to the experimental sample, allowing managers to perform efficient risk mitigation.

\subsubsection{Aggregation}
One naive, but powerful, approach to sequentially test for $\tau_{n,t}(w_{1:n, 1:(t-1)}^{obs})$ is to simply ``stack'' or aggregate the data and apply the same techniques from Section~\ref{section:main_section} to a single time series of $nT$ observations. 

To illustrate this, consider our full panel data matrix
\[
\underbrace{
\begin{bmatrix}
Y_{1, 1} & Y_{1, 2} & \dots & Y_{1, T}  \\
Y_{2, 1} & Y_{2, 2} & \dots & Y_{2, T}\\
\vdots & \vdots & \vdots & \vdots  \\
Y_{n, 1} & Y_{n, 2} & \dots & Y_{n, T} \\
\end{bmatrix}}_{\displaystyle \text{Time}}
\left.\vphantom{\begin{bmatrix}
Y_{1, 1} & Y_{1, 2} & \dots & Y_{1, T}  \\
Y_{2, 1} & Y_{2, 2} & \dots & Y_{2, T}\\
\vdots & \vdots & \vdots & \vdots  \\
Y_{n, 1} & Y_{n, 2} & \dots & Y_{n, T} \\
\end{bmatrix}}\right\} \text{Units} \]
where the rows denote time horizon and the columns denote the $n$ units. At each time $t$, we observe the $t^\text{th}$ column of the above matrix. The naive aggregation method concatenates all columns from the above matrix into one row and models the data as a single time series
$$(\underbrace{Y_{1,1}, \dots, Y_{n,1}}_{t = 1}, \underbrace{Y_{1, 2}, \dots, Y_{n, 2}}_{t = 2}, \dots, \underbrace{Y_{1, T}, \dots, Y_{n, T}}_{t = T}).$$
With this aggregation, we observe $\tilde{t} = 1, 2, \dots, nT$ time points and apply Theorem~\ref{theorem:main_general_thm} on the above aggregated time series. Because at each time $t$ we simultaneously observe $n$ observations, the width of the confidence sequence further decreases with $n$. Since many organizations typically have large number of experimental units, the confidence sequence width can be small even at very early times $t$ and the asymptotics more credible. To formalize this, we further denote the aggregated causal estimates and variance estimates as
$$\hat\tau_{\bullet, t \mid t-1} = \frac{1}{n} \sum_{i = 1}^n \hat\tau_{i, t \mid t-1}, \qquad \hat\sigma_{\bullet, t \mid t-1}^2 = \sum_{i = 1}^n \hat\sigma_{i, t \mid t-1}^2,$$
respectively, where
\begin{align*}
\hat\tau_{i, t \mid t-1} & \coloneqq \frac{\mathbbm{1}\{W_{i,t} = 1 \} Y_{i,t }}{p_{i,t \mid t-1}(1)} - \frac{\mathbbm{1}\{W_{i,t} = 0 \} Y_{i,t}}{p_{i,t \mid t-1}(0)}, \quad \hat\sigma_{i,t \mid t-1}^2  \coloneqq \frac{ \mathbbm{1}\{W_{i,t} = 1 \} Y_{i,t}^2}{p_{i,t \mid t-1}(1)^2} + \frac{\mathbbm{1}\{W_{i,t} = 0 \}Y_{i,t}^2}{p_{i,t \mid t-1}(0)^2} 
\end{align*}
are the corresponding individual level counterparts of $\hat\tau_{t \mid t-1}$ and $\hat\sigma_{t \mid t-1}$, respectively. Finally, we also have the corresponding assumption for Assumption~\ref{assumption:var_no_vanish}.

\begin{assumption}[None Vanishing Variance for Panel Data Setting]
\label{assumption:var_no_vanish_panel}
Let $\text{Var}(\hat\tau_{i, t \mid t-1} \mid \mathcal{F}_{n, t-1}) \leq \sigma_{i, t \mid t-1}^2$, where $\sigma_{i, t \mid t-1}^2$ is equivalent to $\sigma_{t \mid t-1}^2$ (defined in Assumption~\ref{assumption:var_no_vanish}) except it is defined for individual level potential outcomes. Then we assume that
$$\frac{1}{\sum_{j = 1}^t \sigma_{\bullet, j \mid j-1}^2} = o(1) \iff \tilde{S}_{\bullet, t \mid t-1}\coloneqq \sum_{j = 1}^t \sigma_{\bullet, j \mid j-1}^2 \xrightarrow{t \rightarrow \infty} \infty \text{ almost surely,}$$
where $\sigma_{\bullet, t \mid t-1}^2 = \sum_{i = 1}^n \sigma_{i, t \mid t-1}^2$.
\end{assumption}
\begin{theorem}[Design-Based Confidence Sequence for the Average Contemporaneous Treatment Effect for Panel Data]
Suppose $\{W_{i,t}, Y_{i,t}\}_{i,t}$ are observed for all units $i = 1, 2, \dots, n$ and $t = 1, 2, \dots, T$ for any arbitrary data dependent stopping time $T$, where Assumptions~\ref{assumption:ind_PO}-\ref{assumption:var_no_vanish_panel} are satisfied. Denote $\bar{\tau}_{n, t}(w_{1:n, 1:(t-1)}^{obs}) \coloneqq \frac{1}{t}\sum_{j = 1}^t \tau_{n,j}(w_{1:n, 1:(j-1)}^{obs})$ as the running mean for the average contemporaneous treatment effect defined in Equation~\eqref{eq:ATE} and $S_{\bullet, t \mid t-1} \coloneqq \sum_{j = 1}^t  \hat\sigma_{\bullet, j \mid j -1}^2 $. Then,
$$\frac{1}{t} \sum_{j = 1}^t \hat\tau_{\bullet, j \mid j-1} \pm  \frac{1}{tn} \sqrt{ \frac{S_{\bullet, t \mid t-1} \eta^2 + 1}{\eta^2} \log \Bigg( \frac{S_{\bullet, t \mid t-1} \eta^2 + 1}{\alpha^2}\Bigg) } $$
forms a valid $(1-\alpha)$ asymptotic confidence sequence for $\bar{\tau}_{n, t}(w_{1:n, 1:(t-1)}^{obs})$ with approximation rate $\break o\left(\sqrt{\tilde{S}_{\bullet, t \mid t-1} \log \tilde{S}_{\bullet, t \mid t-1}} /tn\right)$ and for any pre-specified constant $\eta > 0$.
\label{theom:panel_aggregation}
\end{theorem}
The proof is omitted because under the stated assumptions it is identical to that of Theorem~\ref{theorem:main_general_thm} replacing time with $\tilde{t} = 1, 2, \dots, nT$ and stacking the panel data into one time series. Theorem~\ref{theom:panel_aggregation} shows that our confidence sequence width further scale approximately with $1/\sqrt{n}$. 

One advantage of the aggregation approach is that it can also account for units that enter the experiment at different times. Although for clarity we present the results when there are exactly the same $n$ units at each time $t$, our method also accommodates staggered adoption when the total number of units $n$ are different at each time $t$, \emph{\emph{i.e.}}, units can both be entering and leaving at any time point $t$. In such a case the causal estimand will also consequently change according to the different units that enter and exit. We present an example using Theorem~\ref{theom:panel_aggregation} in Example~\ref{example:panel_data}.

\section{Variance reduction technique via proxy outcomes}
\label{subsection:proxy_outcome}
In the previous sections, our confidence sequences were constructed with estimators that only leverage the observed treatment and outcomes. In practice, there may be available covariates $X$ or prior information that an analyst can leverage to further reduce the confidence sequence width through the use of proxy outcomes.


\subsection{Proxy outcomes in single time series experiment}
\label{subsection:proxy_outcome_singleTS}
We first illustrate how to incorporate covariates in the general case for the single time series experiment (Section~\ref{section:TS_panel}). We first denote $\hat f_t(X_{1:T}, Y_{1:(t-1)}, W_{1:(t-1)})$ to be the prediction for $Y_t$ using any available covariate information $X_{1:T} = (X_1, X_2, \dots, X_T)$ (each $X_t$ may be multi-dimensional) and our previous data $(W, Y)$ until time $t- 1$. For brevity, we define $\hat Y_{t \mid t-1} \coloneqq \hat f_t(X_{1:T}, Y_{1:(t-1)}, W_{1:(t-1)})$. Since our prediction is based only on a function of our filtration $\mathcal{F}_{X, t-1}$, defined by enriching $\mathcal{F}_{t-1}$ to include the information set for all covariate $X_{1: T}$, \citep{timeseries} defines $\hat Y_{t \mid t-1}$ as a ``time series proxy outcome''. 

The prediction $\hat Y_{t \mid t-1}$ and the filtration are functions of \textit{all} covariate information $X_{1:T}$, including covariate information not necessarily available at time $t < T$. We present it this way to clarify two points. First, in the panel data setting, where the experiment involves $n$ units, the covariate information for all the units may be known beforehand. In such cases, conditioning the covariates for all members at any time $t$ accurately reflects the aforementioned scenario. Second, there may be cases where the covariates change over time or the experiment's units are not initially known, such as in new member experiments. In such situations, the analyst may still want to use $X_t$ to create the prediction $\hat Y_t$ for $Y_t$ To allow this, we allow both the prediction and the filtration to condition on all $X_t$, including the future, similar to how we always condition on all the potential outcomes. In practice, however, the analyst will typically only leverage the available covariates, $X_{1:t}$, for predicting $Y_t$ at time $t$.

Using this proxy outcome, our causal estimand in Definition~\ref{def:causal_estimand} can be rewritten as,
$$\tau_t(w_{1:(t-1)}^{obs}) = \{Y_t(w_{1:(t-1)}^{obs}, 1) - \hat Y_{t \mid t-1} \}  - \{Y_t(w_{1:(t-1)}^{obs}, 0) - \hat Y_{t \mid t-1} \}.$$
Using the above formulation, the corresponding estimator using the proxy outcome is
\begin{equation}
\hat{\tau}_{t \mid t-1}^X \coloneqq \frac{\mathbbm{1}\{W_t = 1 \} \{Y_t - \hat Y_{t \mid t-1} \}}{p_{t \mid t-1}(1)} - \frac{\mathbbm{1}\{W_t = 0 \}\{Y_t - \hat Y_{t \mid t-1} \}}{p_{t \mid t-1}(0)},
\label{eq:proxy_outcome}
\end{equation}
with a similar upper-bound estimate of the variance of
\begin{equation}
    \hat \gamma_{t \mid t-1}^2 \coloneqq \frac{ \mathbbm{1}\{W_t = 1 \} \{Y_t - \hat Y_{t \mid t-1} \}^2}{p_{t \mid t-1}(1)^2} + \frac{\mathbbm{1}\{W_t = 0 \}\{Y_t - \hat Y_{t \mid t-1} \}^2}{p_{t \mid t-1}(0)^2},
\label{eq:sigma2_estimate}
\end{equation}
respectively. One can directly see that $\hat{\tau}_{t \mid t-1}^X$ is again unbiased for $\tau_t(w_{1:(t-1)}^{obs})$ conditional on $\mathcal{F}_{X, t-1}$, where conditioning on all the $X_{1:T}$ does not harm our inference because $\hat Y_{t \mid t-1}$ still remains a constant conditional on the filtration (hence the predictions are constructed only using past data without using current $W_t$). This allows the analyst to formally use the proxy outcome $\hat Y_{t \mid t-1}$ to incorporate any machine learning algorithm or prior knowledge to reduce the variance. This reduction is proportional to how small $\{Y_t - \hat Y_{t \mid t-1} \}^2$ is, \emph{\emph{i.e.}}, how well the analyst can use the prior data to predict the next response. 

Furthermore, we can also allow treatment assignment probabilities $p_{t \mid t-1}(w)$ to depend on the covariates since we always condition on $\mathcal{F}_{X, t-1}$. Lastly, we also require that the new variances with the proxy outcome do not disappear and that the proxy outcomes do not output infinity so that our new outcome $Y_t - \hat Y_{t \mid t-1}$ remains bounded.
\begin{assumption}[Non-Vanishing Variance with Proxy Outcomes]
$\text{Var}(\hat\tau_{t \mid t-1}^X \mid \mathcal{F}_{X, t-1}) \leq \gamma_{t \mid t-1}^2 $, where $\gamma_{t \mid t-1}^2$ is provided in Appendix~\ref{appendix:proof_moments} Equation~\eqref{eq:gamma_exp_appendix} and similar to the expression in Equation~\eqref{eq:sigma_exp} except we replace each potential outcome in the numerator with $Y_t(w_{1:(t-1)}^{obs}, \bullet) - \hat Y_{t \mid t-1}$. Then we assume that
$$\frac{1}{\sum_{j = 1}^t \gamma_{j \mid j-1}^2} = o(1) \iff \tilde{S}_{t \mid t - 1}^X \coloneqq \sum_{j = 1}^t \gamma_{j \mid j-1}^2 \xrightarrow{t \rightarrow \infty} \infty \text{ almost surely.}$$
Further, we have that the predictions do not return infinity, \emph{\emph{i.e.}}, $|\hat Y_{t \mid t-1}| \leq M'$ for all $t$ and $M' \in \mathbb{R}$.
\label{assumption:proxy_var_vanish}
\end{assumption}
\begin{theorem}[Design-based Asymptotic Confidence Sequence Using Proxy Outcomes]
Suppose $\{W_t, Y_t, X_t\}_{t = 1}^T$ are observed for arbitrary data dependent stopping time $T$, where Assumptions~\ref{assumption:PTA}\footnote{In Assumption~\ref{assumption:PTA}, we allow our adaptive probability treatment assignments to adapt to the covariate values $X$ by replacing $\mathcal{F}_{t-1}$ with $\mathcal{F}_{X, t-1}$ for this theorem.}, \ref{assumption:boundedPO_general}, and \ref{assumption:proxy_var_vanish} are satisfied. Let $S_{t \mid t-1}^X \coloneqq \sum_{j = 1}^t \hat\gamma_{j \mid j -1}^2$. Then,
$$ \frac{1}{t} \sum_{j = 1}^t \hat\tau_{j \mid j -1}^X \pm  \frac{1}{t} \sqrt{ \frac{S_{t \mid t-1}^X \eta^2 + 1}{\eta^2} \log \Bigg( \frac{S_{t \mid t-1}^X \eta^2 + 1}{\alpha^2}\Bigg) }$$
forms a valid $(1-\alpha)$ asymptotic confidence sequence for the running mean of the contemporaneous treatment effect $\bar{\tau}_t(w_{1:(t-1)}^{obs})$ with approximation rate $o\left(\sqrt{\tilde{S}_{t \mid t-1}^X \log \tilde{S}_{t \mid t-1}^X}/t\right)$ for any pre-specified constant $\eta > 0$.
\label{theom:proxy_outcome}
\end{theorem}
The proof is omitted because the setting as well as the assumptions are identical to that of Theorem~\ref{theorem:main_general_thm} except we replace $S_{t \mid t-1}$ with $S_{t \mid t-1}^X$ and introducing $\hat Y_{t \mid t-1}^X$ is equivalent to changing the (fixed) potential outcome $Y_t(w_{1:(t-1)}^{obs}, \bullet)$ to a new constant $Y_t(w_{1:(t-1)}^{obs}, \bullet) - \hat Y_{t \mid t-1}^X$. Since the entire proof always conditions on the filtration, the proof remains identical. However, this difference allows the confidence sequence in Theorem~\ref{theom:proxy_outcome} to incorporate covariates and other prior information to potentially reduce the confidence sequence width. Furthermore, our covariates $X_t$ can contain both pre-treatment covariates that do not evolve over time, \emph{\emph{i.e.}}, user sex and race, browser and device type, etc., and time varying covariates that can evolve over time (even as a function of previous treatment assignments making it a post-treatment confounder). Additionally, even if there are no available covariates, one could still likely reduce variance by using the sample mean of $Y_{1:(t-1)}$ for $\hat Y_{t \mid t-1}$, thus making Theorem~\ref{theom:proxy_outcome} a useful practical extension for many cases. 

Although we present Theorem~\ref{theom:proxy_outcome} in the general time series setting with carryover effects, the same variance reduction technique also extends to non-time series setting results we presented in Section~\ref{subsection:n_iid_setting}. In the aforementioned setting, Theorem~\ref{theom:proxy_outcome} would be identical except all expressions with subscript $t$ are replaced with $n$, allowing the proxy outcome $\hat f_{n \mid n-1}$ to instead predict the next unit's response as a function of all the previous data and available covariate information.

\subsection{Proxy outcomes in panel data setting}
\label{subsection:proxyoutcome_panel}
The variance reduction technique via proxy outcomes is most applicable in the panel data setting because we can leverage common information shared across $n$ units (as opposed to only one unit) to make predictions for the next time point. Although generalizing the results in Section~\ref{subsection:proxy_outcome_singleTS} to the panel data setting is straightforward, we formalize this for completeness. 

Starting in a similar fashion, we can rewrite the average contemporaneous treatment effect as
$$\tau_{n,t}(w_{1:n, 1:(t-1)}^{obs}) = \Big\{\frac{1}{n} \sum_{i = 1}^{n} (Y_{i, t}(w_{i, 1:(t-1)}^{obs}, 1) - \hat Y_{i, t \mid t-1} ) \Big\} - \Big\{\frac{1}{n} \sum_{i = 1}^{n}(Y_{i,t}(w_{i, 1:(t-1)}^{obs}, 0) - \hat Y_{i, t \mid t-1} )  \Big\},$$
where $\hat Y_{i, t \mid t-1}$ is the prediction for the $i^\text{th}$ individual's outcome at time $t$ as a function of $\{W_{i, 1:(t-1)}, X_{i, 1:T}, Y_{i, 1:(t-1)}\}_{i = 1}^n$ and $X_{i, t}$ is the (multivariate) covariate value(s) for individual $i$ at time $t$. We denote $\mathcal{F}_{X, n, t-1}$ as the sigma algebra containing $\{W_{i, 1:(t-1)}, X_{i, 1:T}, Y_{i, 1:(t-1)}\}_{i = 1}^n$ and all potential outcomes for all $n$ units up to time $T$. 

For the panel data setting, the corresponding estimators using the proxy outcome are 
$$\hat\tau_{\bullet, t \mid t-1}^X = \frac{1}{n} \sum_{i = 1}^n \hat\tau_{i, t \mid t-1}^X, \qquad \hat\gamma_{\bullet, t \mid t-1}^2 = \sum_{i = 1}^n \hat\gamma_{i, t \mid t-1}^2,$$
respectively, where
\begin{align*}
\hat\tau_{i, t \mid t-1 }^X & \coloneqq \frac{\mathbbm{1}\{W_{i,t} = 1 \} \{Y_{i,t} - \hat Y_{i, t \mid t-1}\} }{p_{i,t \mid t-1}(1)} - \frac{\mathbbm{1}\{W_{i,t} = 0 \} \{Y_{i,t} - \hat Y_{i, t \mid t-1}\} }{p_{i,t \mid t-1}(0)} \\
\hat\gamma_{i,t \mid t-1}^2 & \coloneqq \frac{ \mathbbm{1}\{W_{i,t} = 1 \} \{Y_{i,t} - \hat Y_{i, t \mid t-1}\}^2}{p_{i,t \mid t-1}(1)^2} + \frac{\mathbbm{1}\{W_{i,t} = 0 \}\{Y_{i,t} - \hat Y_{i, t \mid t-1}\}^2}{p_{i,t \mid t-1}(0)^2}.
\end{align*}
Similar to Assumption~\ref{assumption:proxy_var_vanish}, we have the following assumption.
\begin{assumption}[Non-Vanishing Variance with Proxy Outcome Variances for Panel Data]
Let $\text{Var}(\hat\tau_{i, t \mid t-1}^X \mid \mathcal{F}_{X, n, t-1}) \leq \gamma_{i, t \mid t-1}^2 $, where $\gamma_{i, t \mid t-1}^2$ is equivalent to $\gamma_{t \mid t-1}^2$ (defined in Assumption~\ref{assumption:proxy_var_vanish}) except for individual level potential outcomes and adaptive probability assignments. Then we assume that
$$\frac{1}{\sum_{j = 1}^t \gamma_{\bullet, j \mid j-1}^2} = o(1) \iff \tilde{S}_{\bullet, t \mid t - 1}^X \coloneqq \sum_{j = 1}^t \gamma_{\bullet, j \mid j-1}^2 \xrightarrow{t \rightarrow \infty} \infty \text{ almost surely,}$$
where $\gamma_{\bullet, t \mid t-1}^2 = \sum_{i = 1}^n \gamma_{i, t \mid t-1}^2$. Further, we have that the predictions do not return infinity, \emph{i.e.}, $|\hat Y_{i, t \mid t-1}| \leq M'$ for all $i, t$ and $M' \in \mathbb{R}$.
\label{assumption:proxy_var_vanish_panel}
\end{assumption}
\begin{theorem}[Design-based Asymptotic Confidence Sequence Using Proxy Outcomes for Panel Data]
Suppose $\{W_{i,t}, Y_{i,t}, X_{i,t}\}_{i, t}$ are observed for arbitrary data dependent stopping time $T$ and $i = 1, 2, \dots, n$ units, where Assumptions~\ref{assumption:ind_PO}, \ref{assumption:panel_data_regularity}\footnote{In Assumption~\ref{assumption:panel_data_regularity}, we further allow our adaptive probability treatment assignments to adapt to covariate values $X$ by replacing $\mathcal{F}_{n, t-1}$ with $\mathcal{F}_{X, n, t-1}$ for this theorem.}, and \ref{assumption:proxy_var_vanish_panel} are satisfied. Let $S_{\bullet, t \mid t-1}^X \coloneqq \sum_{j = 1}^t \hat\gamma_{\bullet, j \mid j -1}^2$. Then,
$$ \frac{1}{t} \sum_{j = 1}^t \hat\tau_{\bullet, j \mid j -1}^X \pm  \frac{1}{tn} \sqrt{ \frac{S_{\bullet, t \mid t-1}^X \eta^2 + 1}{\eta^2} \log \Bigg( \frac{S_{\bullet, t \mid t-1}^X \eta^2 + 1}{\alpha^2}\Bigg) } $$
forms a valid $(1-\alpha)$ asymptotic confidence sequence for the running mean of the average contemporaneous treatment effect $\bar{\tau}_{n, t}(w_{1:n, 1:(t-1)}^{obs})$ with approximation rate $o\left(\sqrt{\tilde{S}_{\bullet, t \mid t-1}^X \log \tilde{S}_{\bullet, t \mid t-1}^X}/t\right)$ for any pre-specified constant $\eta > 0$.
\label{theom:proxy_outcome_paneldata}
\end{theorem}
Comparing Theorem~\ref{theom:proxy_outcome_paneldata} and Theorem~\ref{theom:panel_aggregation}, we see that we can again get a reduction in variance depending on how small we can make $\{Y_{i,t} - \hat Y_{i, t \mid t-1}\}^2$, \emph{i.e.}, how well we can predict the next response for each individuals at every time using past data. We end this section with an example demonstrating Theorem~\ref{theom:panel_aggregation} and the consequent reduction in variance we can get using Theorem~\ref{theom:proxy_outcome_paneldata}.

\begin{figure}[t]
\begin{center}
\includegraphics[width=15cm]{"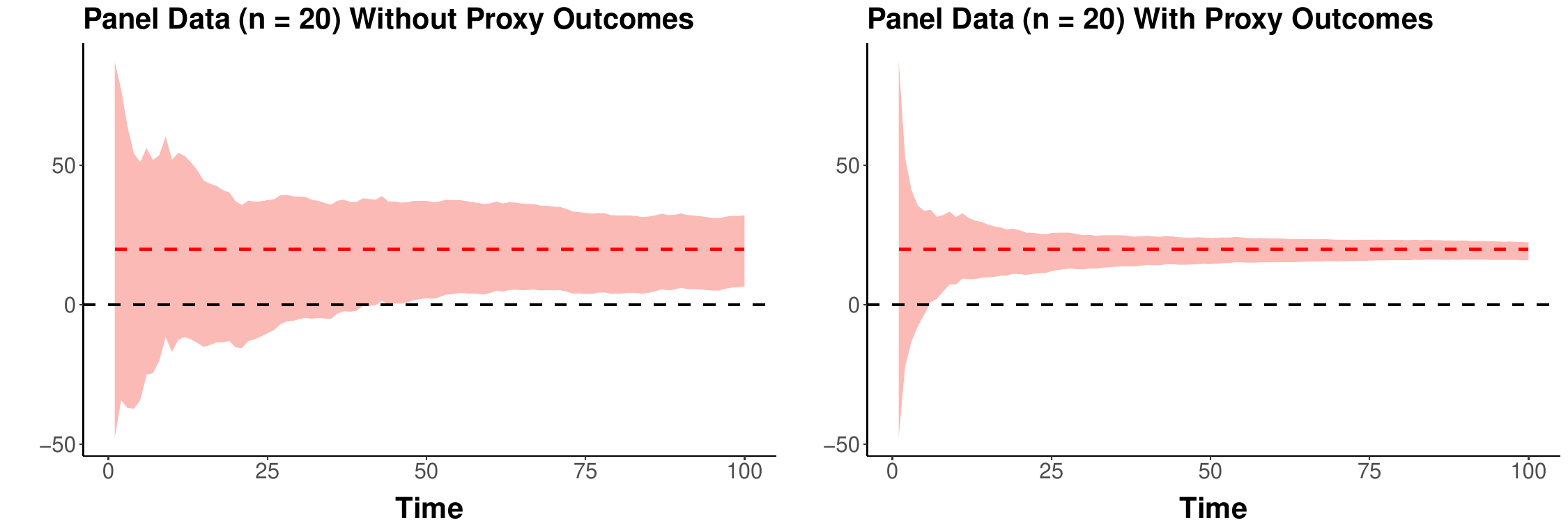"}
\caption{Unit Varying Treatment Effects in Linear Models (Example~\ref{example:panel_data}). The red contours show the confidence sequence for the average treatment effect for the 20 individuals in our sample at $\alpha = 0.05$ using Theorem~\ref{theom:panel_aggregation} (left panel) and Theorem~\ref{theom:proxy_outcome_paneldata} (right panel). The parameters for this example are $n = 20, T = 100, \beta = 1, \rho = 0.5, \mu = 20$. The red dotted line represents the true average treatment effect that diminishes to zero and the black horizontal dotted line represents the zero (null) line. }
\label{fig:panel_data_ex}
\end{center}
\end{figure}

\begin{example}[Unit Varying Treatment Effects in Linear Models]
\label{example:panel_data}
Suppose that every individual has an engagement score that is a function of the previous day's engagement and covariate $X$. Further suppose that each user has a user-specific treatment effect. More formally, 
\begin{align*}
    Y_{i,t}(0) &= \rho Y_{i, t-1}(0) + \beta X_i + \epsilon_{i, t}, \quad |\rho| \leq 1, \quad \epsilon_{i,t} \overset{iid}{\sim} N(0, 10^2) \\
    Y_{i, t}(1) &= Y_{i, t}(0) + \mu_i, \quad \mu_i \sim N(\mu, 10^2) \\
    Y_{i, 0} & \overset{iid}{\sim} \beta X_i + \epsilon_{i, 0}, \quad X_i \overset{iid}{\sim} N(25, 5^2), \quad  W_{i, t} \overset{iid}{\sim} \text{Bern}(0.5) \text{ for all } i, t
\end{align*}
This scenario reflects a practical example where each unit may react differently to the treatment, but overall the treatment effect changes the engagement level by $\mu$ on average. For simplicity, we use a time-invariant covariate $X_i$ that also has a stationary relationship $\beta$ across time. Lastly, all examples thus far have assumed the potential outcomes come from some independent distribution. Because the design-based approach conditions on the potential outcome, we can allow for any arbitrary dependence. In this case, we demonstrate it for an AR(1) process, where each user's current engagement is only dependent on previous engagement. We set $n = 20, T = 100, \beta = 1, \rho = 0.5, \mu = 20$ for this example.

For the left panel of Figure~\ref{fig:panel_data_ex}, we build the confidence sequence using Theorem~\ref{theom:panel_aggregation} by pretending we do not observe $X_i$. For the right panel of Figure~\ref{fig:panel_data_ex}, we build the confidence sequence using Theorem~\ref{theom:proxy_outcome_paneldata}, where $\hat Y_{i, t \mid t-1}^X$ is the predicted response from an ordinary least square regression of all responses $Y$ on all covariates $X$ available at time $t$. Consequently, as $t$ grows, $\hat\beta$ becomes more accurate, allowing a reduction in variance proportional to $(\beta X_i)^2$. Figure~\ref{fig:panel_data_ex} shows that using proxy outcomes substantially reduces the confidence sequence width. For example, the right panel would reject the null average treatment effect by $t = 6$ while the left panel would reject it by $t = 43$, approximately a seven times reduction. 
\end{example}

\section{Simulation study}
\label{section:simulations}
\subsection{\dwh{Robustness of asymptotic results}}
\label{subsection:sim_robustness}
As Examples~\ref{example:carryover} and \ref{example:panel_data} each construct only a single CS, they do not confirm whether our proposed confidence sequences have the time-uniform type-1 error guarantee. \dwh{Furthermore, as Theorem~\ref{theorem:main_general_thm} is an asymptotic result, the desired \textit{uniform} type-1 error guarantees may not hold for finite samples. In this section, we show the validity and robustness of our asymptotic results in finite samples. 

To achieve this, we replicate the simulation setting in Example~\ref{example:panel_data} except simplify the scenario to exclude any covariates $X$. We include $X$ when demonstrating how to reduce the variance through proxy outcomes in the next  Section~\ref{subsection:sim_1}.  Removing $X$, we now have the following data-generating process,
$$Y_{t}(0) = \rho Y_{t-1}(0) + \epsilon_t, \quad Y_t(1) = Y_t(0) + \mu, \quad W_t \overset{iid}{\sim} \text{Bern}(0.5),$$
where we drop subscript $i$ and focus on the single time series setting for simplicity. We set $\rho = 0.5$ and obtain the treatment effect $\mu \overset{iid}{\sim} N(1, 0.5^2)$, where we replicate this simulation for 5000 empirical Monte-Carlo runs. We control type-1 error at level $\alpha = 0.05$. 

To demonstrate the robustness of our asymptotic approximations, we increase the difficulty for the asymptotics to take effect by varying the distribution of the noise, $\epsilon_t$, under three settings. 1) We make $\epsilon_t \overset{iid}{\sim} N(0, 1)$ such that our outcomes are already exactly Gaussian in finite-samples. 2) We make $\epsilon_t \overset{iid}{\sim} 1 - \text{Exp}(1)$ from the standard exponential distribution. The exponential distribution has a strong right skew, making the Gaussian asymptotic approximations take longer to be valid. 3) We make $\epsilon_t \overset{iid}{\sim} \text{Cauchy}$ from the standard Cauchy distribution. The Cauchy distribution poses issues for asymptotic approximations since the Cauchy distribution does not have well defined moments. 

For each Monte-Carlo simulation, we report two time-dependent measures of type-1 error. First, we report the proportion of times the confidence sequence in Theorem~\ref{theorem:main_general_thm} failed to cover the true treatment effect $\mu$ at every time $t$ in the left plot of Figure~\ref{fig:robustness_fig}. For example, the green curve peaked at around $0.020$ at $t = 4$, showing that the confidence sequence in Theorem~\ref{theorem:main_general_thm} failed to capture the true treatment effect $\mu$ approximately 2\% of the times at time $t = 4$ when the errors $\epsilon_t$ were distributed from a standard exponential distribution. We remark that the uniform type-1 error is not simply the addition of the type-1 errors across all time $t$ in the left plot of Figure~\ref{fig:robustness_fig}. For example, if the confidence sequence failed to cover the truth at $t = 4$, it very likely failed to the cover the truth also at $t = 3$, thus many of these type-1 errors are overlapped. 

To also report the uniform type-1 error guarantee by time, we plot in the right panel of Figure~\ref{fig:robustness_fig} the cumulative type-1 error had one started checking at time $t$. For example, the point at $t = 1$ represents the uniform type-1 error had the analyst checked the confidence sequence at every time starting from the first data point $t = 1$. The point at $t = 5$ represents the uniform type-1 error had the analyst not ``peeked'' for the first 4 samples $t = 1, 2, 3, 4$ but started peeking at every sample from $t = 5, 6, \dots, T$ . Since the cumulative type-1 error at $t = 1$ is, by definition, the uniform type-1 error had the analyst started checking right away (including all early times), we further show this uniform type-1 error with horizontal dotted lines. 

\begin{figure}[t]
\begin{center}
\includegraphics[width=14cm]{"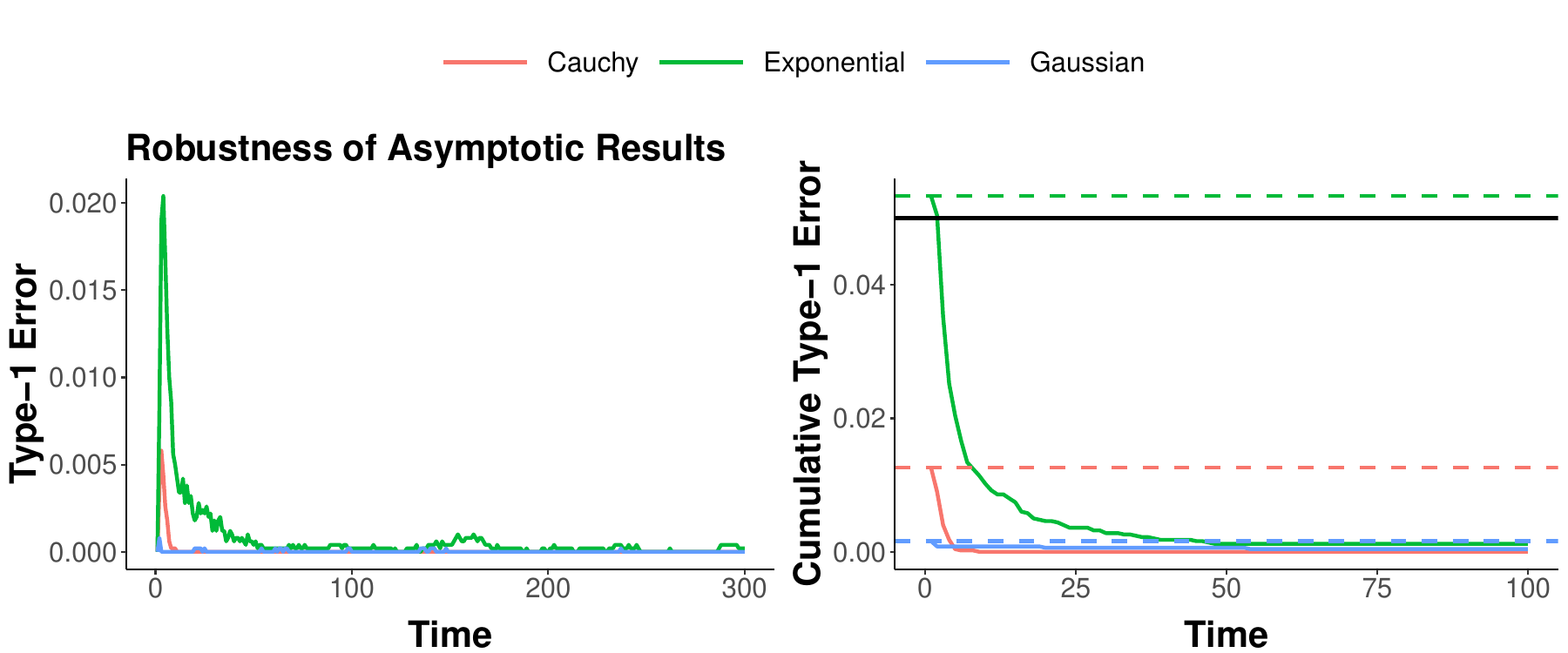"}
\caption{\dwh{Demonstrating type-1 error validity for Theorem~\ref{theorem:main_general_thm}. The left panel shows the proportion of times the confidence sequence in Theorem~\ref{theorem:main_general_thm} at time $t$ failed to capture the truth. The right panel shows the cumulative uniform type-1 error  had the analyst started peeking from time $t$. The horizontal lines represent the uniform type-1 error had the analyst started peeking right away from the first data point $t = 1$. The simulation setting is identical to that in Example~\ref{example:panel_data} without any $X$ and only one unit $n = 1$. We vary $\epsilon_t$ to be generated $iid$ from a Gaussian (blue curve), standard Exponential (green curve), and standard Cauchy (red curve) distribution. We set $\rho = 0.5, \alpha = 0.05$. The simulation results were over 5000 empirical Monte-Carlo runs.}}
\label{fig:robustness_fig}
\end{center}
\end{figure}

Figure~\ref{fig:robustness_fig} shows that our asymptotic results are robust in finite samples even with skewed data. The over conservative coverage is due to our variance estimate being an upper bound estimate. For example, when the noise distribution is Gaussian, our type-1 error is almost zero for all time points (blue curve). Although the Cauchy and Exponential distributed errors results in more miscoverage at earlier times, the uniform type-1 errors, $0.053, 0.013$, is still roughly $\alpha$ or lower than $\alpha$ for the Exponential (green dotted line) and Cauchy distributed errors (blue dotted lines), respectively. Although the Exponential distributed errors cause the empirical uniform coverage to be close to $\alpha$, the left panel of Figure~\ref{fig:robustness_fig} shows the majority of miscoverage is happening early when attempting to build an asymptotic-CS with just a few samples. However, the right panel of Figure~\ref{fig:robustness_fig} shows that if the analyst started checking after only $t = 5$ samples, the uniform type-1 error drops from $0.053$ to $0.020$ for the Exponential distributed errors. Therefore, we find that our proposed asymptotic confidence sequences are valid even in small times and robust to skewed or non-Gaussian data as similarly shown in \citep{bibaut, time_uniform}. Although, our simulation results show that one could directly just check as early as the first data-point, we recommend analysts to start ``peeking'' after some initial time, $t \geq 5$, to allow the asymptotics to take effect and avoid making type-1 errors in the very beginning. We believe this is also good practice as analysts should not typically make conclusive inference based off only two to five data points. 
}

\subsection{\dwh{Variance reduction via proxy outcomes}}
\label{subsection:sim_1}

In this section, we bring back covariates $X$ and demonstrate variance reduction through proxy outcomes (Theorem~\ref{theom:proxy_outcome}). To achieve this, we first replicate the simulation in Example~\ref{example:panel_data} for $5000$ empirically computed confidence sequences and record the proportion of times it contained the true treatment effect for all times $t$ and the average time it took to reject the point null of zero treatment effect, which we refer to as ``average stopping time.'' For the second scenario, we adjust the linear model with the following nonlinear model
\begin{equation}
Y_{i,t}(0) = \rho Y_{i, t-1}(0) + |x \sin(x)| + \epsilon_{i, t}
\label{eq:non_linear_setup}
\end{equation}
while using the same OLS prediction so that the prediction model is misspecified. 


\begin{table}[!t]
\begin{center}
\begin{adjustbox}{max width=\textwidth,center}
\begin{tabular}{l|cc|} 
& \multicolumn{2}{c|}{Average Empirical Results (5000 MC Repetitions)}  \\
Different Scenarios & Type-1 Error & Stopping Time  \\ 
 \hline
Scenario 1: Without proxy outcome (linear model: $n = 20$) & 0.002 & 36 \\
Scenario 1: With well specified proxy Outcome (linear model: $n = 20$) & 0.002 & 5.5 \\
Scenario 2: Without proxy outcome (non-linear model: $n = 20$) & 0.001 & 34 \\
Scenario 2: With misspecified proxy outcome (non-linear model: $n = 20$) & 0.001 & 29 
\end{tabular}
\end{adjustbox}
\caption{The first two rows represent simulations under the same setting as that in Example~\ref{example:panel_data}. The third and fourth rows are also under the same setting except we change the potential outcome model to Equation~\eqref{eq:non_linear_setup}. The last row shows the empirical type-1 error control when $n = 1$ under scenario one. The second column represents the empirical proportion of times each respective confidence sequence covers the true treatment effect for all times $t$ when $\alpha = 0.05$. The third column represents the average time it took to reject the point null of zero treatment effect. The proportion and average are taken over 5000 Monte-Carlo simulated confidence sequences for each scenario.}
\label{tab:mainresults}
\end{center}
\end{table}

Table~\ref{tab:mainresults} show the results of the simulation. As expected, scenarios 1-2 show strong type-1 error control even at early times. Further, we see approximately a seven times reduction in the average stopping time using proxy outcomes under scenario 1. Although the difference is not as substantial when the prediction model is misspecified for a highly non-linear outcome, we still roughly see a 15\% decrease in the average stopping time. 


\subsection{Comparison to confidence intervals and hybrid approaches}
\label{subsection:sim2}
We next aim to quantify the tradeoffs practitioners face when deciding whether to perform anytime-valid inference (confidence sequence), fixed-time standard inference (confidence interval), or a mixture of both. To facilitate this comparison, we  compare confidence intervals and a naive interim hybrid approach under a similar simulation setting as above.

To construct valid confidence intervals, we use the design-based asymptotically valid confidence interval proposed in \citep{timeseries}. We choose this comparison because it is centered around the same inverse propensity score estimator and uses the same design-based variance estimator in Equation~\eqref{eq:sigma_estimate}, allowing a fair comparison. Alternatively, practitioners may opt for a hybrid approach that is the middle ground of confidence intervals and confidence sequences such as group sequential test that conducts hypothesis testing for only a finite number of times \citep{GST, GST_book}. This strategy uses multiple-testing correction procedures to achieve the uniform type-1 error control in Equation~\eqref{eq:validity} for finite number of times. To compare with this method, we use a Bonferroni-corrected confidence interval from \citep{timeseries} and construct $K$ confidence intervals at uniform times between $t = 1, \dots, T$ as suggested by \citep{GST}. For example, if $T = 190$ and the hybrid approach tests at $K = 5$ uniformly spaced points, then we construct a Bonferroni-corrected confidence interval at $t = 38, 76, 114, 152, 190$. 

For our simulation, we again simplify the data generating process in Example~\ref{example:panel_data}. First we set $\rho = \beta = 0$ to get rid of any structural dependency and covariates. Next, we set the treatment mean $\mu = 10$ and set $n = 2$. \dwh{To decide the final stopping time $T$ for the confidence interval approach, we do a power calculation to achieve $90\%$ power. We do this to reflect a more practical approach analysts take when constructing confidence intervals.  We calculate this empirically and obtain $T = 190$ to achieve 90\% power.} We report both the type-1 error and the average stopping time as similarly done in Table~\ref{tab:mainresults}. However, to facilitate a fair comparison of ``stopping time'' with confidence intervals and the hybrid interim approach, we redefine stopping time as the fastest time to detect a positively significant effect assuming the experiment is always terminated at $T = 190$ (even if a positive effect is not detected). Since confidence intervals are tested only once at $T = 190$, the stopping time for confidence intervals is trivially 190. The stopping time for confidence sequences is the average times it stopped early before $t < 190$ and averaging all the times it stopped at $t = 190$ because it was unable to find a statistically significant effect. 

We additionally report both the average width and statistical power. The average width is defined as the expected width of the confidence interval, sequence, or hybrid confidence interval at the final time $T = 190$. The statistical power is the probability of rejecting the null treatment effect by time $t = T$. For example, if the statistical power of the confidence sequence is 80\%, then by $t = T$ the confidence sequence on average rejects the null treatment effect with a 80\% chance before the end of the experiment.

\begin{table}[!t]
\begin{center}
\begin{adjustbox}{max width=\textwidth,center}
\begin{tabular}{l|cccc|}
& \multicolumn{4}{c|}{Average Empirical Results (5000 MC Repetitions)}  \\
Method & Type-1 Error  & Stopping Time &  Width & Power \\ 
\hline
Confidence Interval & 98\% & 190 &  5.61 & 90\% \\
Hybrid Interim Approach  & 97\% & 74.0 & 7.36 & 88\% \\
Confidence Sequence & 96\% & 70.3 & 11.56  & 79\% \\
\end{tabular}
\end{adjustbox}
\caption{Simulations comparing confidence sequence and intervals under the setting in Example~\ref{example:panel_data} with $\rho = \beta = 0, \mu = 10, T = 190, n = 2$. The first row shows the performance of asymptotically valid design-based confidence intervals proposed in \citep{timeseries}. The second row performs a hybrid interim approach by constructing the aforementioned valid confidence interval at five uniform finite times $t = 38, 76, 114, 152, 190$ and performing a test at each time. The last row constructs the confidence sequence using Theorem~\ref{theom:proxy_outcome} that is valid for all times. The second and third column are defined similarly as the simulation in Table~\ref{tab:mainresults}. The last two columns denote the average width at $T = 190$ and the statistical power of each method, respectively.}
\label{tab:sim_CI_comparison}
\end{center}
\end{table}

Table~\ref{tab:sim_CI_comparison} shows that the confidence interval, as expected, has a higher statistical power than the confidence sequence. However, the confidence sequence, on average, can terminate much earlier by $t = 70$ rather than $t = 190$, reducing the total expected experimentation time by roughly 65\%. While confidence sequences have a wider interval than confidence intervals by definition, we quantify this tradeoff by demonstrating that the confidence sequence width is penalized by approximately a factor of two compared to the width of a confidence interval. Finally, the hybrid interim approach ($K = 5$) performs worse than both the confidence interval in terms of power and width, but it may have the potential to stop earlier than confidence intervals while still having competitive stopping times with confidence sequences. Although this approach provides a suitable middle ground between the two extremes, the advantages of potentially stopping early is sensitive to the pre-specified finite-times. Furthermore, the analyst can not continue to run the experiment even if the analyst desires to after the final pre-determined time, posing similar issues to confidence intervals.  

\section{Application to Netflix experiments}
\label{section:application}

\dwh{Netflix has a strong experimentation culture, leveraging A/B test to inform decisions on product development, provide safety checks on software deployments, and more \citep{michael_paper, michael_paper2}. Anytime-valid approaches to analyzing tests are of interest to Netflix for two reasons. First, as a risk mitigation technique so that deleterious treatments can be rapidly  identified and shut off. Second, as a way to increase innovation velocity by ending successful experiments early, so that all members can more rapidly benefit from the innovation and Netflix teams can move onto the next idea. In the latter case, the exchange of power for earlier expected stopping times is of particular interest. Here we reanalyze three experiment run by Netflix to explore how the design based confidence sequences would have permitted for more more rapid decision making. }

\subsection{Messaging treatment}
\label{subsection:messaging}
We now reanalyze an experiment that tested whether or not sending a push messaging notification (for example, the notification could remind subscribers that a specific show has been released) increases engagement among Netflix members. To do this, we leverage Theorem~\ref{theom:panel_aggregation} to build a confidence sequence due to the special panel data structure.

\begin{figure}[t]
\begin{center}
\includegraphics[width=12cm]{"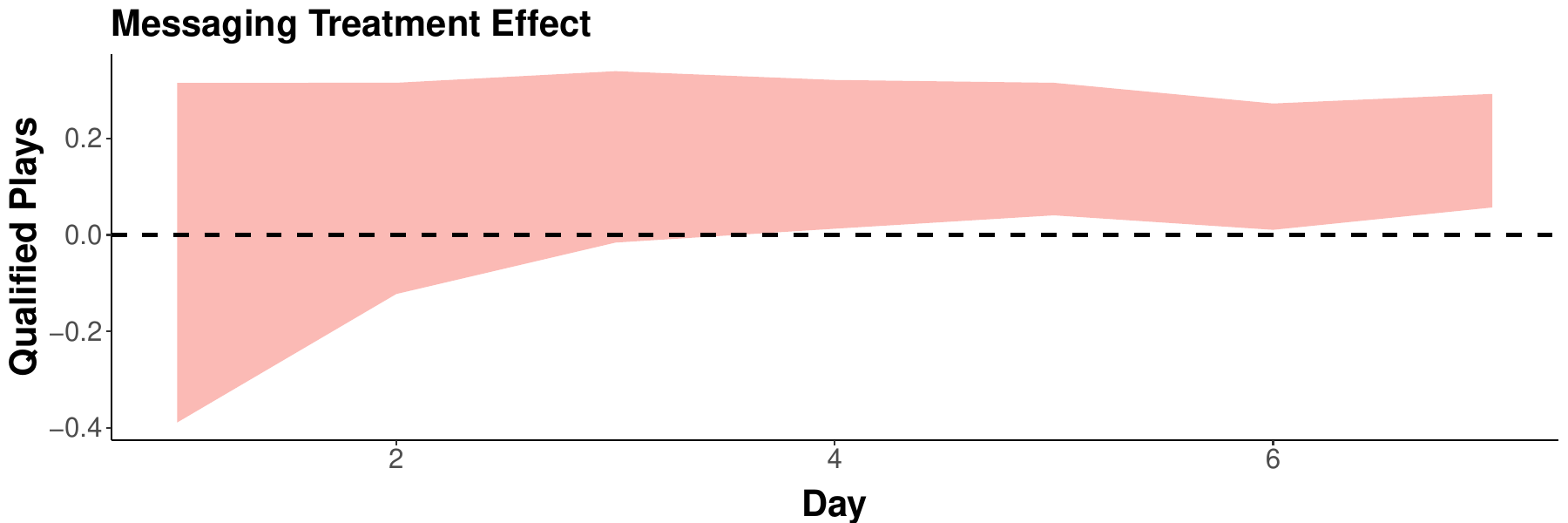"}
\caption{Messaging treatment application. The red contours show the lower and upper confidence sequence as approximately 2000 units continue to see messages or not throughout approximately one week at $\alpha = 0.05$ using Theorem~\ref{theom:panel_aggregation}. The black horizontal dotted line represents the zero (null) line.}
\label{fig:app2}
\end{center}
\end{figure}

Netflix ran the experiment on $n \approx 2000$ members that have some probability of seeing a message or not (our binary treatment) on each day $t$. The probabilities were computed from a machine learning model (details omitted) that predicts whether each member would benefit from a message. Finally, the experiment was run for approximately one week, where it is known that sending messages this way has a strong positive impact on engagement. The primary response of interest is ``Qualified Play'', an internal metrics that records if a user has played a video for a meaningful amount of time on that day. Figure~\ref{fig:app2} shows that even with only $n \approx 2000$ users, we see a positive treatment effect as early as the fourth day with a final estimated treatment effect of $0.17$. This allows the manager to terminate the experiment earlier in this more complex panel data setting. 

\subsection{Sign-up page experiments}
\label{subsection:prepaid}
A crucial component of any subscription-based business is designing a simple and efficient sign-up page that attracts new members. Consequently, Netflix runs thousands of experiments to adapt and improve the sign-up page while attempting to mitigate the experimental risk of losing potential members. In this section, we analyze two such experiments. 


The first experiment tests whether providing users with verification and assistance when entering credit card information can help more prospective members complete the sign-up process. Netflix has observed that users often benefit from assistance when completing their subscriptions after incorrectly inputting credit card information--- a well-known phenomenon studied in UX research \citep{UX_research}. To improve the process, Netflix designed a new interface that automatically detects the brand of the card (\emph{\emph{e.g.}}, Mastercard or Visa) after the first few numbers and automatically formats the card numbers (\emph{\emph{e.g.}}, XXXX-XXXX-XXXX-XXXX) corresponding to the brand. To test the effectiveness of the new design, Netflix ran an experiment in a single country, comparing the new design against the standard offering without such assistance; we refer to this experiment/treatment as the ``Credit Card Assistance'' experiment/treatment. 

Over approximately\footnote{For the purposes of this paper, 
 we omit the exact details of the data, such as precise sample sizes and test durations.} two weeks, $N \approx 30,000$ visitors to the Netflix sign up page were randomly assigned (with equal probability) to either the treatment (the new sign-up page) or the control (the original sign-up page). The primary outcome was whether or not the individual successfully joined Netflix by providing a working payment method (binary outcome). 

The second experiment tests whether clearly indicating that Netflix accepts prepaid gift cards drove additional sign-ups. Currently, Netflix accepts prepaid gift cards, but this is not indicated in the payment selection. Managers at Netflix believed that indicating this information clearly would help more prospective members join Netflix; we refer to this experiment as the ``Prepaid Card'' experiment. 

The Prepaid Card experiment was conducted for approximately 2 months in Italy with a sample size of $N \approx 125, 000$ total visitors to the Netflix sign up page that were either shown the default experience (control) or a different version where Netflix clearly indicates that they accept Postepay cards, a popular Italian prepaid card (treatment). \dwh{Netflix used a top-two Thompson adaptive sampling scheme as they believed that rapidly moving towards the best treatment would help maximize subscriptions \citep{thompson}. The top-two Thompson sampling scheme assures that the two promising arms always have some positive probability of being treated, immediately satisfying Assumption~\ref{assumption:PTA}. For this reason, we also focus on estimating the treatment effect between these two top arms. }

\begin{figure}[t]
\begin{center}
\includegraphics[width=12cm]{"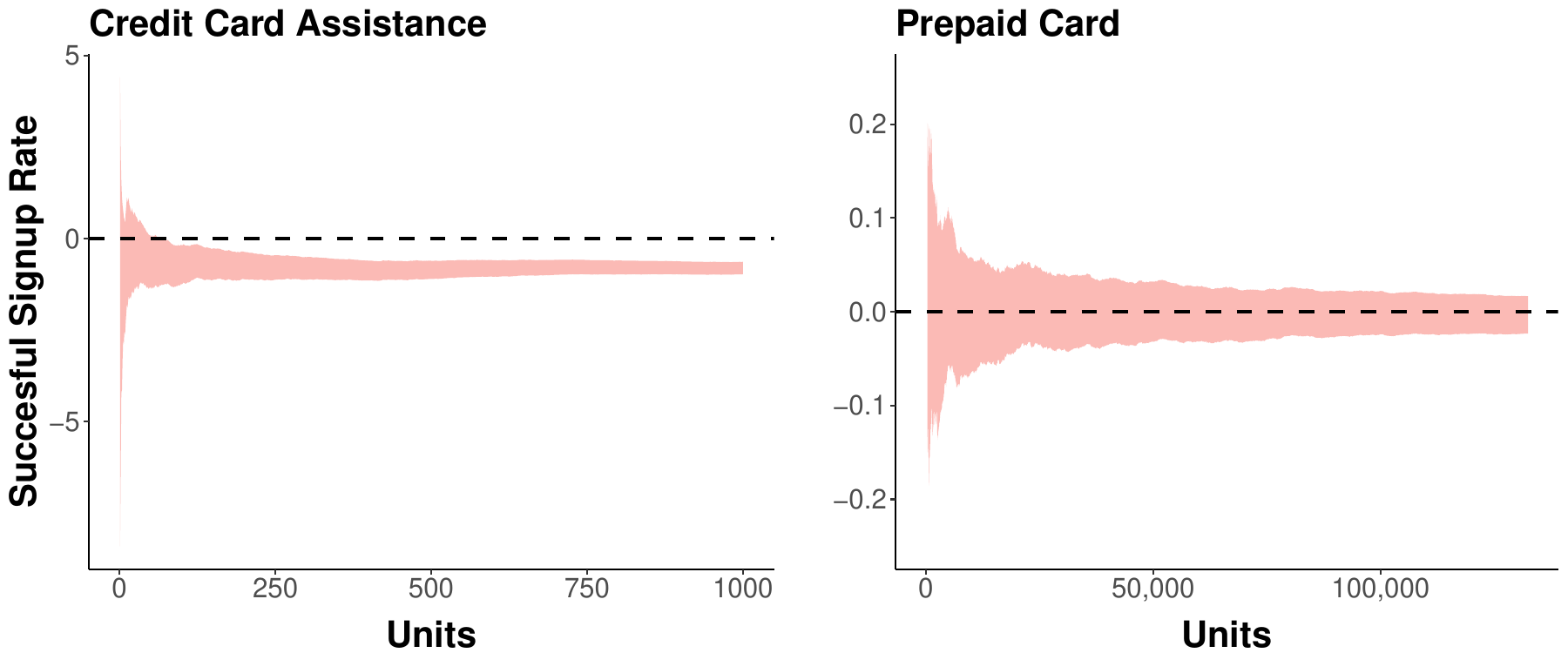"}
\caption{Sign-up page experiments. The left panel shows the confidence sequence using Theorem~\ref{theorem:main_general_thm} for the Credit Card Assistance experiment. We plot the confidence sequence for the first $n \approx 1000$ units out of a total sample of $N \approx 30,000$. The right panel shows the confidence sequence to analyze the adaptive Prepaid Card experiment. The black horizontal dotted line represents the zero (null) line with $\alpha = 0.05$. }
\label{fig:app1}
\end{center}
\end{figure}

We re-analyze the two sign-up page experiments in Figure~\ref{fig:app1} using Lemma~\ref{theom:main_thm1_ATE} for both sign-up experiments. The left panel plots the confidence sequence using only the first $n = 1000$ customers out of a total sample size of $N \approx 30,000$. We see that the confidence sequence detects a statistically significant negative treatment effect before the $70^\text{th}$ unit and a treatment effect at least as harmful as $5\%$ by approximately the $80^\text{th}$ unit. This result is consistent with the finding at Netflix, where this experiment was known to have a negative treatment effect due to technical issues related to implementing the treatment for the Firefox browser. However, this experiment ran for approximately two weeks with this harmful treatment on over $N = 30,000$ customers with a final estimated treatment effect of approximately $-20\%$\footnote{The point estimate is an estimate for the average treatment effect $\tau_N$, \emph{i.e.}, $\sum_{i = 1}^N \tau_i/N \approx -0.20$}. The left panel of Figure~\ref{fig:app1} shows this could have been avoided had the business manager used confidence sequences and terminated the harmful experiment as early as the first day before incurring further costs. 

The right panel shows the confidence sequence when analyzing the Prepaid Card experiment. We find that the confidence sequence covers zero for all units, showing that indicating that pre-paid cards are accepted does not lead to higher conversion rates with a point estimate of $-0.0033$. This result is also consistent with other findings, where Netflix ran another non-adaptive experiment for the same setting and found no significant treatment effect using a $t$-test with over $100,000$ samples. Suppose the analyst wanted to terminate the experiment as soon as there is evidence the treatment effect is less than $\epsilon$. In that case, Figure~\ref{fig:app1} shows the analyst would terminate the experiment approximately as early as the $60,000^\text{th}$ unit for $\epsilon = 0.05$, \emph{i.e.}, the first unit where the analyst can confidently conclude the treatment effect is less than 5\%. For $\epsilon = 0.02$, the analyst would approximately terminate at the $100,000^\text{th}$ unit. 

\section{Concluding remarks}
\label{section:discussion}

Our work bridges the sequential testing literature with the design-based literature to perform continuous monitoring for the contemporaneous treatment effect relevant also to the bandit settings and the average contemporaneous treatment effect in panel data settings with time-uniform guarantees. 

Our confidence sequences formally allow managers to ``peek'' at any time and stop the experiment in a data-dependent way, \emph{\emph{e.g.}}, as soon as detecting a statistically significant harmful effect. \dwh{Through the design-based approach, our confidence sequences are valid for a larger, previously unstudied, class of experiments in time series and panel data settings with minimal assumptions.} Additionally, the design-based approach allows managers to perform risk mitigation by quantifying an estimand directly relevant to the harm incurred on the obtained experimental sample.


\section*{Acknowledgement}
We thank Ian Waudby-Smith and participants of the 2022 Conference on Digital Experimentation for advice and feedback.
\newpage
\putbib 
\end{bibunit}

\newpage
\section*{Online Appendix \\ Design-Based Confidence Sequences: A General Approach to Risk
Mitigation in Panel Experiments}
\subsection*{Dae Woong Ham, Michael Lindon, Martin Tingley, and Iavor Bojinov}
\appendix

\begin{bibunit}

\section{Example for multi-arm bandit}
\label{appendix:n_iid_setting}

We first remark that our propensity scores in Lemma~\ref{theom:main_thm1_ATE} for the $n$ units, $p_{i \mid i-1}(w)$, can be fully adaptive. This now allows us to tackle the common bandit settings, where an agent is typically tasked to maximize reward and pull the next arm as a function of all the previous treatment assignments \citep{MAB_overview, RLbook}. Although the positivity assumption in Assumption~\ref{assumption:PTA} for time series and panel experiments is more likely to hold and controlled by the experiment, in adaptive bandit settings Assumption~\ref{assumption:PTA} may not hold asymptotically if the adaptive algorithm converges to pulling an optimal arm with probability one. Therefore, this assumption is often a ``stricter'' assumption for multi-arm bandits as many adaptive algorithms such as Thompson Sampling \citep{thompson, top_two} make the adaptive probabilities converge to one or zero. To solve this concern, we recommend practitioners to use a mixture of a uniform explore policy and an adaptive exploit policy, \emph{\emph{e.g.}}, assigning an epsilon mixture weight to the uniform explore policy ensures the propensity scores are bounded. We now show an example of using Lemma~\ref{theom:main_thm1_ATE} for a simplified two arm bandit setting. 

\begin{figure}[b]
\begin{center}
\includegraphics[width=12cm]{"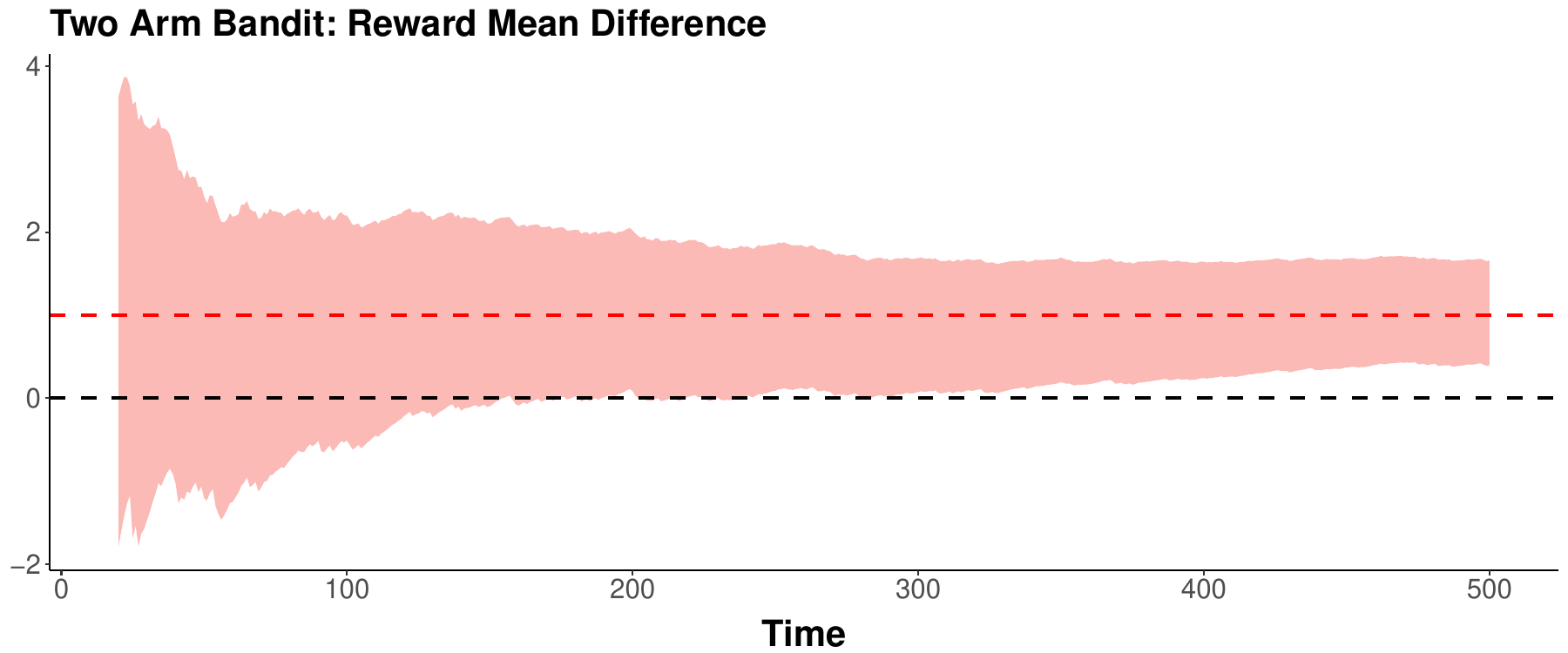"}
\caption{Two Arm Bandit (Example~\ref{example:bandit}). The red contours show the lower and upper confidence sequence as the agent pulls each arm adaptively at $\alpha = 0.05$ using Lemma~\ref{theom:main_thm1_ATE} . The horizontal red dotted line represent the true mean difference of the rewards and the black horizontal dotted line represents the zero (null) line. }
\label{fig:bandit_example}
\end{center}
\end{figure}

\begin{example}[Two Arm Bandit]
\label{example:bandit}
Suppose for simplicity the two-arm bandit problem, where an agent pulls either arm A (control) or arm B (treatment). Suppose the rewards under arms A and B have distributions $N(1, 1)$ and $N(2,1)$, respectively. Consider the following adaptive probabilities for the $n^\text{th}$ individual 
$$p_{n \mid n-1}(1) = \frac{\bar{Y}_{1, n-1}}{\bar{Y}_{1, n-1} + \bar{Y}_{0, n-1}}, \quad n > 10$$
and $p_{n \mid n-1}(0) = 1-  p_{n \mid n-1}(1)$, where $\bar{Y}_{1, n-1}$ is the sample mean of arm B using the realized samples for $i = 1, 2, \dots, n-1$ and $\bar{Y}_{0, n-1}$ is defined similarly. In other words, the agent upweights the arm that produce a higher mean reward based on the sample means. Lastly, the agent flips a fair coin for the first ten time periods (exploration period). Figure~\ref{fig:bandit_example} shows that even under adaptively sampled data, our confidence sequence tightens to the desired truth. For this case, the agent would likely terminate the experiment at approximately $n = 150$ because the reward from arm $B$ is statistically significantly higher than the reward from arm $A$. 
\end{example}

\section{Proof of Theorem~\ref{lemma:moment_cond}}
\label{appendix:proof_moments}

$$E(\hat\tau_t \mid \mathcal{F}_{t-1}) = \tau_t(w_{1:(t-1)}^{obs})$$
\begin{proof}
\begin{align*}
  E(\hat\tau_t \mid \mathcal{F}_{t-1}) &= E\left(\frac{\mathbbm{1}\{W_t = 1 \} Y_t}{p_{t \mid t-1}(1)} - \frac{\mathbbm{1}\{W_t = 0 \} Y_t}{p_{t \mid t-1}(0)} \mid \mathcal{F}_{t-1} \right) \\
 &= \frac{p_{t \mid t-1}(1) Y_t(w_{1:(t-1)}^{obs}, 1) }{p_{t \mid t-1}(1)} - \frac{ p_{t \mid t-1}(0) Y_t(w_{1:(t-1)}^{obs}, 0)}{p_{t \mid t-1}(0)}  \\
 &= \tau_t(w_{1:(t-1)}^{obs}).
\end{align*}
\end{proof}

Next, we calculate the closed form expression of $\text{Var}(\hat\tau_{t \mid t-1} \mid \mathcal{F}_{t-1})$.
\begin{align*}
    \text{Var}(\hat\tau_{t \mid t-1} \mid \mathcal{F}_{t-1}) &= \text{Var}\left(\frac{\mathbbm{1}\{W_t = 1 \} Y_t}{p_{t \mid t-1}(1)} - \frac{\mathbbm{1}\{W_t = 0 \} Y_t}{p_{t \mid t-1}(0)} \mid \mathcal{F}_{t-1} \right) \\
    &= \frac{p_{t \mid t-1}(1)p_{t \mid t-1}(0)  Y_t(w_{1:(t-1)}^{obs}, 1)^2}{p_{t \mid t-1}(1)^2} + \frac{p_{t \mid t-1}(0)p_{t \mid t-1}(1) Y_t(w_{1:(t-1)}^{obs}, 0)^2}{p_{t \mid t-1}(0)^2} \\
    &+ 2Y_t(w_{1:(t-1)}^{obs}, 1)Y_t(w_{1:(t-1)}^{obs}, 0)  \\
    &= \frac{p_{t \mid t-1}(0)  Y_t(w_{1:(t-1)}^{obs}, 1)^2}{p_{t \mid t-1}(1)} + \frac{p_{t \mid t-1}(1) Y_t(w_{1:(t-1)}^{obs}, 0)^2}{p_{t \mid t-1}(0)} + 2Y_t(w_{1:(t-1)}^{obs}, 1)Y_t(w_{1:(t-1)}^{obs}, 0)  \\
    &= \frac{\left(p_{t \mid t-1}(0)  Y_t(w_{1:(t-1)}^{obs}, 1) + p_{t \mid t-1}(1)  Y_t(w_{1:(t-1)}^{obs}, 0) \right)^2}{p_{t \mid t-1}(1)p_{t \mid t-1}(0)},
\end{align*}
where the third line follows because $\text{Cov}(\mathbbm{1}\{W_t = 1 \}, \mathbbm{1}\{W_t = 0 \} \mid \mathcal{F}_{t-1}) = - p_{t \mid t-1}(1) p_{t \mid t-1}(0)$. Now we show that
$$\text{Var}(\hat\tau_{t \mid t-1} \mid \mathcal{F}_{t-1})  \leq \sigma_{t \mid t-1} \coloneqq \frac{Y_t(w_{1:(t-1)}^{obs}, 1)^2}{p_{t \mid t-1}(1)} + \frac{Y_t(w_{1:(t-1)}^{obs}, 0)^2}{p_{t \mid t-1}(0)},$$
which would complete the proof because it is straight forward to show that $E(\hat\sigma_{t \mid t-1} \mid \mathcal{F}_{t-1}) = \sigma_{t \mid t-1}$. 
\begin{proof}
\begin{align*}
\text{Var}(\hat\tau_{t \mid t-1} \mid \mathcal{F}_{t-1}) & = \frac{p_{t \mid t-1}(0)^2  Y_t(w_{1:(t-1)}^{obs}, 1)^2 + p_{t \mid t-1}(1)^2  Y_t(w_{1:(t-1)}^{obs}, 0)^2 }{p_{t \mid t-1}(1)p_{t \mid t-1}(0)} \\   
& + \frac{2 p_{t \mid t-1}(0)  Y_t(w_{1:(t-1)}^{obs}, 1) p_{t \mid t-1}(1)  Y_t(w_{1:(t-1)}^{obs}, 0)}{p_{t \mid t-1}(1)p_{t \mid t-1}(0)} \\
& \leq \frac{Y_t(w_{1:(t-1)}^{obs}, 1)^2}{p_{t \mid t-1}(1)} + \frac{Y_t(w_{1:(t-1)}^{obs}, 0)^2}{p_{t \mid t-1}(0)},
\end{align*}
\noindent where the last line follows because $(a - b)^2 \geq 0 \rightarrow a^2 + b^2 \geq 2ab$ and we let $a = p_{t \mid t-1}(0)  Y_t(w_{1:(t-1)}^{obs}, 1)$ and $b = p_{t \mid t-1}(1)  Y_t(w_{1:(t-1)}^{obs}, 0)$.
\end{proof}

Lastly, $\gamma_{t \mid t-1}^2$ defined in Assumption~\ref{assumption:proxy_var_vanish} can be obtained by replacing $Y_t(w_{1:(t-1)}^{obs}, \bullet)$ with a new ``residualized'' potential outcome $\tilde{Y}_t(w_{1:(t-1)}^{obs}, \bullet) := Y_t(w_{1:(t-1)}^{obs}, \bullet) - \hat Y_{t \mid t-1}$. Since all the proof is conditioned on the filtration, the above is equivalent to introducing a constant and therefore we have that
\begin{equation}
\label{eq:gamma_exp_appendix}
 \gamma_{t \mid t-1}^2 =  \frac{\tilde{Y}_t(w_{1:(t-1)}^{obs}, 1)^2}{p_{t \mid t-1}(1)} + \frac{\tilde{Y}_t(w_{1:(t-1)}^{obs}, 0)^2}{p_{t \mid t-1}(0)}.   
\end{equation}

\section{Alternative closed-form exact design-based confidence sequence}
\label{appendix:proof_nonasymp}
To derive an alternative exact confidence sequence, we apply the empirical Bernstein inequalities to derive a design-based confidence sequence \citep{empirical_bernstein, fan_ineq}. 

\begin{theorem}[Design-based Non-asymptotic Confidence Sequence for the CTE]
\label{theorem:nonasymp_CS}
Suppose $\{W_t, Y_t\}_{t = 1}^T$ are observed for arbitrary data dependent stopping time $T$, where Assumptions~\ref{assumption:PTA} and \ref{assumption:boundedPO_general} are satisfied. Then, 
  $$ \frac{1}{t} \sum_{j = 1}^t \hat\tau_{j \mid j-1} \pm \left[\frac{m(m+1)}{t} \log \Bigg( \frac{2}{\alpha} \Bigg) + \frac{S_{t \mid t-1}}{t} \left(\frac{m+1}{m}\log\Big(1 + \frac{1}{m} \Big) - \frac{1}{m} \right) \right] $$
forms a valid $(1-\alpha)$ confidence sequence for \dwh{the running mean of the contemporaneous average treatment effect $\bar{\tau}_t(w_{1:(t-1)}^{obs})$}.
\end{theorem}
\noindent The proof is provided in the bottom of this Appendix section. The key part of the proof is that we show
$$\exp\left[\frac{\sum_{j = 1}^t\hat\tau_{j \mid j-1} - \tau_t(w_{1:(t-1)}^{obs})}{m(m+1)}  + \frac{S_{t \mid t-1}}{m^2} \left(\log\Big(\frac{m}{m+1} \Big) + \frac{1}{m+1} \right) \right]$$
forms a supermartingale by applying Fan's inequality \citep{fan_lemma} through similar proof techniques in \citep{howard_nonasymp}. The rest follows from algebraically manipulating the supermartingale after applying Lemma~\ref{lemma:ville}.

We note that the confidence width is of order approximately $S_{t \mid t-1} /t$, which does not shrink asymptotically ($t \rightarrow \infty$) to zero unless $S_{t \mid t-1}$ grows sub-linearly ($\hat\sigma_t$ vanishes to zero) or there are stronger assumptions on the potential outcomes. This issue was recognized in \cite{howard_nonasymp} who propose a solution under additional conditions. Therefore, we similarly propose Theorem~\ref{theorem:nonasymp_CS_gamma_mixture} instead. We still present a simulation figure using this exact CS Example~\ref{example:nonasymp} below.

Before stating the proof we quickly show an example and simulation to demonstrate this confidence sequence. 

\begin{figure}[t]
\begin{center}
\includegraphics[width=15cm]{"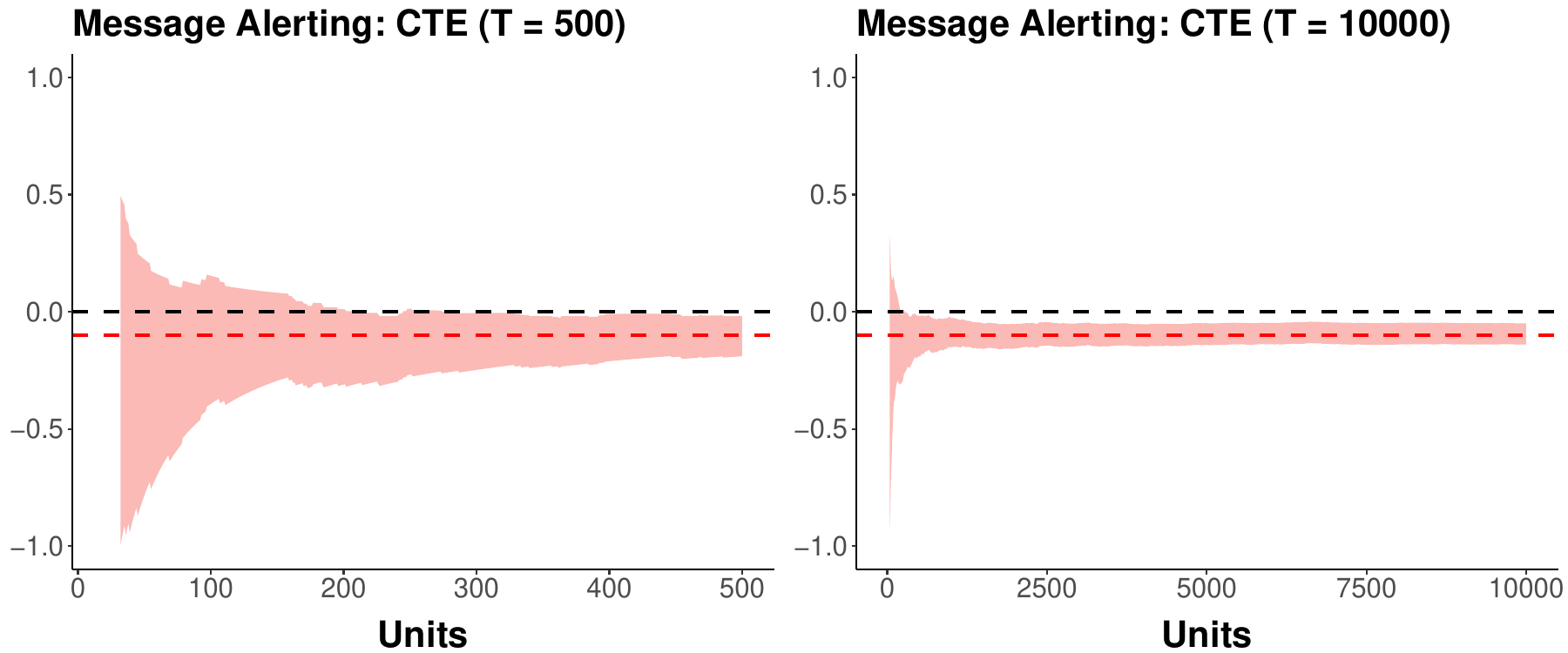"}
\caption{Online Experimentation (Example~\ref{example:nonasymp}). The red contours show the lower and upper confidence sequence as units enter in the experiment at $\alpha = 0.05$ using Theorem~\ref{theorem:nonasymp_CS}. The left and right panel show the confidence sequences for $T = 500, 10000$ units, respectively. The horizontal red dotted line represent the true treatment effect of $\tau_n = 0.1$ and the black horizontal dotted line represents the zero (null) line. }
\label{fig:nonasymp_example}
\end{center}
\end{figure}

\begin{example}[Simple Messaging Alerting Treatment]
\label{example:nonasymp}
Suppose the same scenario as Example~\ref{example:carryover} but simplified to a stationary constant treatment effect, \emph{i.e.}, no novelty effect but there is a constant treatment effect of $0.10$ across time. 
Then, in the notation from Theorem \ref{theorem:nonasymp_CS}, $M = 1$ and $m = 1/0.5 = 2$. Furthermore, the \dwh{the running mean of the CTE} is roughly -0.1 at any time $t$. \dwh{Although this scenario is largely simplified to remove any non-stationary, i.e., the user responds exactly the same way to a message alert at any time $t$, we present it this way to pedagogically demonstrate the theoretical properties of Theorem~\ref{theorem:nonasymp_CS}.}
The left and right panels of Figure~\ref{fig:nonasymp_example} show the confidence sequence from applying Theorem~\ref{theorem:nonasymp_CS} on one simulated experimental data in the aforementioned setting for $T = 500$ and $T = 10,000$ time points, respectively. The left panel shows that the analyst would likely stop at the $t = 210^\text{th}$ time because the confidence sequence shows a statistically significant negative treatment effect (the first time the red contours do not overlap the black dotted line), mitigating further harm. However, the right panel shows that the confidence width does not shrink despite a large $T = 10,000$, illustrating the aforementioned limitations. 
\end{example}

We now prove Theorem~\ref{theorem:nonasymp_CS}. 
We build off the proof of Theorem 4 in \citep{howard_nonasymp} and apply it to our specific time series setting. As hinted in Section~\ref{subsection:nonasympc_CS}, we will first show that 
$$\exp\left[\frac{\sum_{j = 1}^t\hat\tau_{j \mid j-1} - \tau_t(w_{1:(t-1)}^{obs})}{m(m+1)}  + \frac{S_{t \mid t-1}}{m^2} \left(\log\Big(\frac{m}{m+1} \Big) + \frac{1}{m+1} \right) \right]$$
is a non-negative supermartingale with respect to the filtration $\mathcal{F}_{N, n-1}$. \citep{fan_lemma} show that 
$$\exp\left( \lambda\kappa  + \kappa^2 (\lambda + \log(1 - \lambda)) \right) \leq 1 + \lambda \kappa$$
for $\kappa \geq -1$ and $\lambda \in [0, 1)$ from Fan's Inequality. We let 
$$\kappa = \frac{\hat\tau_{t \mid t-1}}{m},$$
where $\kappa \geq -1$ since $|\hat \tau_{t \mid t-1}| \leq m$ for every $t$ by Assumption~\ref{assumption:boundedPO_general}. Therefore, we have 
\begin{align*}
\exp\left( \lambda \frac{ \hat\tau_{t \mid t-1}}{m}  +  \frac{\hat\tau_{t \mid t-1}^2}{m^2}(\lambda + \log(1 - \lambda)) \right) & \leq 1 + \lambda \frac{ \hat\tau_{t \mid t-1}}{m} \\
E \left[\exp\left( \frac{\lambda \hat\tau_{t \mid t-1}}{m}  + \frac{\hat\sigma_{t \mid t-1}^2}{m^2} (\lambda + \log(1 - \lambda)) \right) \mid \mathcal{F}_{t-1} \right] & \leq 1 + \lambda \frac{\tau_t(w_{1:(t-1}^{obs})}{m} \\
E \left[\exp\left( \frac{\lambda (\hat\tau_{t \mid t-1} - \tau_t(w_{1:(t-1}^{obs}))}{m}  + \frac{\hat\sigma_{t \mid t-1}^2}{m^2} (\lambda + \log(1 - \lambda)) \right) \mid \mathcal{F}_{t-1} \right] & \leq \exp(-\frac{\lambda \tau_t(w_{1:(t-1}^{obs})}{m})[1 + \lambda \frac{\tau_t(w_{1:(t-1}^{obs})}{m}] \\
E \left[\exp\left( \frac{\lambda (\hat\tau_{t \mid t-1} - \tau_t(w_{1:(t-1}^{obs}))}{m}  + \frac{\hat\sigma_{t \mid t-1}^2}{m^2} (\lambda + \log(1 - \lambda)) \right) \mid \mathcal{F}_{t-1} \right] & \leq 1,
\end{align*}
where the second line follows because $\hat\tau_{t \mid t-1}^2 = \hat\sigma_{t \mid t-1}^2$ and Theorem~\ref{lemma:moment_cond} and the last line follows because $1 + x \leq \exp(x)$. We plug $\lambda = 1/(m+1)$ and because the above is a non-negative quantity. This directly implies that 
$$\exp\left[\frac{\sum_{j = 1}^t \hat\tau_{j \mid j-1} - \tau_t(w_{1:(t-1)}^{obs})}{m(m+1)}  + \frac{S_{t \mid t-1}}{m^2} \left(\log\Big(\frac{m}{m+1} \Big) + \frac{1}{m+1} \right) \right]$$
is indeed a non-negative super martingale with respect to $\mathcal{F}_{N, n-1}$ as desired with initial value less than one. We apply Lemma~\ref{lemma:ville} and have that
\begin{align*}
\Pr\left(\exists t: \exp\left[\frac{\sum_{j = 1}^t(\hat\tau_{j \mid j-1} - \tau_t(w_{1:(t-1)}^{obs}))}{m(m+1)}  + \frac{S_{t \mid t-1}}{m^2} \left(\log\Big(\frac{m}{m+1} \Big) + \frac{1}{m+1} \right) \right] \geq \frac{1}{\tilde{\alpha}} \right) &\leq \tilde{\alpha}  \\
\Pr\left(\exists t: \left[\frac{\sum_{j = 1}^t(\hat\tau_{j \mid j-1} - \tau_t(w_{1:(t-1)}^{obs}))}{m(m+1)}  + \frac{S_{t \mid t-1}}{m^2} \left(\log\Big(\frac{m}{m+1} \Big) + \frac{1}{m+1} \right) \right] \geq \log \Bigg( \frac{1}{\tilde{\alpha}} \Bigg) \right) &\leq \tilde{\alpha}  \\
\Pr\left(\exists t: \sum_{j = 1}^t(\hat\tau_{j \mid j-1} - \tau_t(w_{1:(t-1)}^{obs})) \geq m(m+1) \log \Bigg( \frac{1}{\tilde{\alpha}} \Bigg) -  \frac{(m+1)S_{t \mid t-1}}{m} \left(\log\Big(\frac{m}{m+1} \Big) + \frac{1}{m+1} \right)  \right) &\leq \tilde{\alpha}  \\
\Pr\left(\exists t: \sum_{j = 1}^t(\hat\tau_{j \mid j-1} - \tau_t(w_{1:(t-1)}^{obs})) \geq \left[m(m+1) \log \Bigg( \frac{1}{\tilde{\alpha}} \Bigg) + S_{t \mid t-1} \left(\frac{m+1}{m}\log\Big(1 + \frac{1}{m} \Big) - \frac{1}{m} \right) \right] \right) &\leq \tilde{\alpha}.  \\
\end{align*}

Consequently, we have that
$$\Pr\left(\exists t: \frac{1}{t} \sum_{j = 1}^t(\hat\tau_{j \mid j-1} - \tau_t(w_{1:(t-1)}^{obs})) \geq \left[\frac{m(m+1)}{t} \log \Bigg( \frac{1}{\tilde{\alpha}} \Bigg) + \frac{S_{t \mid t-1}}{t} \left(\frac{m+1}{m}\log\Big(1 + \frac{1}{m} \Big) - \frac{1}{m} \right) \right] \right) \leq \tilde{\alpha}.$$
This gives the one-sided confidence sequence and we can do the same trick and build the same statement instead for $\kappa = -\hat\tau_{t \mid t-1}/m$. Then we get
$$\Pr\left(\exists t: -\frac{1}{t} \sum_{j = 1}^t(\hat\tau_{j \mid j-1} - \tau_t(w_{1:(t-1)}^{obs})) \geq \left[\frac{m(m+1)}{t} \log \Bigg( \frac{1}{\tilde{\alpha}} \Bigg) + \frac{S_{t \mid t-1}}{t} \left(\frac{m+1}{m}\log\Big(1 + \frac{1}{m} \Big) - \frac{1}{m} \right) \right] \right) \leq \tilde{\alpha}.$$
Taking $\alpha = \tilde{\alpha}/2$ and applying the union bound completes the proof. 

\section{Proof of Theorem~\ref{theorem:nonasymp_CS_gamma_mixture}}
\label{appendix:proof_extension}
The above proof shows that
\begin{equation}
\label{eq:super_martingale}
M_t \coloneqq \exp\left( \lambda A_t  + B_t (\lambda + \log(1 - \lambda)) \right) 
\end{equation}
is a super-martingale with initial value 1, where 
$$A_t \coloneqq \frac{\sum_{j = 1}^t \hat \tau_{t \mid t-1} - \tau_t(w_{1:(t-1)}^{obs}) }  {m}, \quad B_t \coloneqq \frac{S_{t \mid t-1}}{m^2}.$$
For any distribution $F$ on $(0,1)$, we have by Fubini's theorem that 
$$\tilde{M}_t \coloneqq \int_{\lambda \in (0, 1)} M_t dF(\lambda) $$
is again another super-martingale with initial value 1. Following the proof of Theorem 2 in \citep{ian_bandit}, we choose the truncated gamma distribution given by
$$f(\lambda) = \frac{\delta^\delta e^{-\delta \left(1 - \lambda\right)} \left(1 - \lambda\right)^{\delta - 1}}{\Gamma(\delta) - \Gamma(\delta, \delta)}$$
for any $\delta \geq 0$. Therefore, we have that 
\begin{align*}
\tilde{M}_t &= \int_0^1 \exp \left \{ \lambda A_t + B_t(\lambda + \log(1 - \lambda) \right \} f(\lambda) d\lambda \\
&=\int_0^1 \exp \left \{ \lambda A_t + B_t(\lambda + \log(1 - \lambda) \right \} \frac{\delta^\delta e^{-\delta \left(1 - \lambda\right)} \left(1 - \lambda\right)^{\delta - 1}}{\Gamma(\delta) - \Gamma(\delta, \delta)} d\lambda \\
&= \frac{\delta^\delta e^{-\delta}}{\Gamma(\delta) - \Gamma(\delta, \delta)} \int_0^1 \exp \{\lambda \left(A_t + B_t + \delta \right)\} \left(1 - \lambda\right)^{B_t + \delta - 1} d\lambda \\
&= \left(\frac{\delta^\delta e^{-\delta}}{\Gamma(\delta) - \Gamma(\delta, \delta)}\right) \left(\frac{1}{B_t + \delta}\right) {_1F_1}(1, B_t + \delta + 1, A_t + B_t + \delta),
\end{align*}
where the last line follows from the definition of the Kummer's confluent hypergeometric function. 

Therefore, we have by Lemma~\ref{lemma:ville} that 
$$\Pr\left(\exists t: \left(\frac{\delta^\delta e^{-\delta}}{\Gamma(\delta) - \Gamma(\delta, \delta)}\right) \left(\frac{1}{B_t + \delta}\right) {_1F_1}(1, B_t + \delta + 1, A_t + B_t + \delta) \geq \frac{1}{\tilde{\alpha}} \right) \leq \tilde{\alpha}.$$
Consequently, a one-sided lower confidence sequence, \emph{i.e.}, a confidence sequence that is bounded from below but covers up to infinity, can be obtained by a root-finding algorithm to find all 
$$\{\tau_t(w_{1:(t-1)}^{obs}): V_t(\tau_t(w_{1:(t-1)}^{obs}) \geq \frac{1}{\tilde{\alpha}} \},$$
where 
$$V_t(\tau_t(w_{1:(t-1)}^{obs})) \coloneqq \left(\frac{\delta^\delta e^{-\delta}}{\Gamma(\delta) - \Gamma(\delta, \delta)}\right) \left(\frac{1}{B_t + \delta}\right) {_1F_1}(1, B_t + \delta + 1, A_t + B_t + \delta).$$
An upper confidence sequence can be obtained in a similar way by using the mirror trick and replacing $\tau_t(w_{1:(t-1)}^{obs})$ with $1 - \tau_t(w_{1:(t-1)}^{obs})$. Furthermore, this confidence sequence does not solve the issue where it requires the analyst to know $M$ and $p_{min}$ before the experiment. Additionally, this confidence sequence does not have a closed-form expression, thus it requires a root-solving algorithm to build the confidence sequence. However, \citep{ian_bandit} show that this provably has an asymptotic rate of $O(\sqrt{B_t \log(B_t)}/t)$, which does solve the issue related to the order of the confidence sequence width. 

\section{Proof of Theorem~\ref{theorem:main_general_thm}}
\label{appendix:proof_main_them}
The proof proceeds in three steps. We note that this proof leverages the proof of Theorem 2.3 in \citep{time_uniform} but extended to our setting. 

\paragraph{Step 1: Building martingale using Gaussian distribution}
Recently, \citep{martingale_must} shows that all sequential tests must have an explicit or implicit construction of a non-negative martingale. Although one of the major advantages of an asymptotic confidence sequences is that it avoids explicitly constructing a martingale, the proof still relies on constructing a martingale with the asymptotic Gaussian distribution. Consequently, the first step of the proof builds a martingale from a sequence of $iid$ standard Gaussian random variables. 

Let $(Z_t)_{t= 1}^{\infty}$ be a sequence of $iid$ standard Gaussian random variable. We note that
$$M_t(\lambda) \coloneqq \text{exp}\left( \sum_{j = 1}^t (\lambda \sigma_{j \mid j-1} Z_j - \lambda^2 \sigma_{j \mid j-1}^2/2) \right)$$
is a non-negative martingale starting at one for any $\lambda \in \mathbb{R}$ with respect to the canonical filtration \citep{filtration_cite}. For algebraic simplicity, we also define $L_t \coloneqq \sum_{j = 1}^t \sigma_{j \mid j-1} Z_j$ and $\bar{\sigma}_t^2 = \frac{1}{t} \sum_{j = 1}^t \sigma_{j \mid j-1}^2$. Moreover, for any probability distribution $F(\lambda)$ on $\mathbb{R}$, we also have the mixture, 
$$\int_{\lambda \in \mathbb{R}} M_t(\lambda) dF(\lambda) $$
is again a non-negative martingale with initial value one \citep{filtration_cite}. In particular, we consider the probability distribution function $f(\lambda; 0, \eta^2)$ for the Gaussian distribution with mean zero and variance $\eta^2$ as the mixing distribution. The resulting martingale is
\begin{align*}
    M_t \coloneqq&  \int_{\lambda \in \mathbb{R}} M_t(\lambda) f(\lambda; 0, \eta^2) d\lambda \\
    =& \frac{1}{\sqrt{2\pi\eta^2}} \int_{\lambda} \text{exp}\left(\lambda  L_t - \frac{t\lambda^2 \bar{\sigma}_t^2 }{2}  \right) \text{exp}\left( \frac{-\lambda^2}{2\eta^2}\right) d\lambda \\
    =& \frac{1}{\sqrt{2\pi\eta^2}} \int_{\lambda} \text{exp}\left(\lambda  L_t - \frac{\lambda^2(1 + t\eta^2 \bar{\sigma}_t^2)}{2\eta^2} \right) d\lambda \\
    =& \frac{1}{\sqrt{2\pi\eta^2}} \int_{\lambda} \text{exp}\left(\frac{-\lambda^2(1 + t\eta^2 \bar{\sigma}_t^2 ) + 2 \lambda \eta^2 L_t} {2\eta^2}  \right) d\lambda \\
    =& \frac{1}{\sqrt{2\pi\eta^2}} \int_{\lambda} \text{exp}\left(\frac{-a(\lambda^2 + \frac{b}{a} 2\lambda} {2\eta^2}  \right) d\lambda ,
\end{align*}
where $a = t\eta^2\bar{\sigma}_t^2 + 1$ and $b = \eta^2 L_t$. Completing the square, we have that the integrand is:
$$\text{exp}\left(\frac{-a(\lambda^2 + \frac{b}{a} 2\lambda} {2\eta^2}  \right)  =  \text{exp}\left(\frac{-(\lambda - b/a)^2}{2\eta^2/a}\right) \text{exp}\left(\frac{b^2}{2a\eta^2}\right).$$
Putting the expression back into $M_t$ we have that,
\begin{align*}
M_t &= \frac{1}{\sqrt{2\pi\eta^2/a}}   \int_{\lambda} \text{exp}\left(\frac{-(\lambda - b/a)^2}{2\eta^2/a}\right) d\lambda \frac{\text{exp}\left(\frac{b^2}{2a\eta^2}\right)}{\sqrt{a}} \\
&= \text{exp}\left( \frac{\eta^2(\sum_{j=1}^t \sigma_{j \mid j-1} Z_j)^2}{2(t \bar{\sigma}_t^2 \eta^2 + 1)}\right) (t \bar{\sigma}_t^2 \eta^2 + 1)^{-1/2},
\end{align*}
\noindent where the last line follows because the first part of the first line is one and we plug back in the definition of $a$ and $b$.

Since $M_t$ is a non-negative martingale with initial value one, we can use Lemma~\ref{lemma:ville} to claim that 
\begin{equation}
\label{eq:gaussian_CS}
\begin{aligned}
&\Pr(\forall t \geq 1, M_t < 1/\alpha) \geq 1 - \alpha \\
=& \Pr\left(\forall t \geq 1, \left|\frac{1}{t} \sum_{j = 1}^t \sigma_{j \mid j-1} Z_j\right| < \sqrt{ \frac{2(t\bar{\sigma}_t^2 \eta^2 + 1)}{t^2 \eta^2} \text{log}\Bigg( \frac{\sqrt{t\bar{\sigma}_t^2\eta^2 + 1}}{\alpha} \Bigg)} \right) \geq 1 - \alpha,
\end{aligned}
\end{equation}
where the last line follows from taking the logarithm and simple algebraic manipulation.

\paragraph{Step 2: Strong Approximation via Martingale Sequence Differences}
We first define the estimation error as
$$u_t = \hat\tau_{t \mid t-1} - \tau_t(w_{1:(t-1)}^{obs}).$$ 
By Theorem~\ref{lemma:moment_cond}, $\{u_t\}$ is a martingale difference sequence with respect to $\mathcal{F}_{t-1}$. Similar to the proof of Step 2 of \citep{time_uniform}, we also use the strong approximation theorem presented in \citep{stratssen}. In particular, we require Equation (159) in Theorem 4.4 of Strassen's paper (further details in Lemma A.3 of \citep{time_uniform}) for our strong approximation theorem. However, our proof is different than that in Step 2 of \citep{time_uniform} for the following reason. 

The original Theorem 4.4 in \citep{stratssen} is stated for martingales difference sequence of the form $E(D_n \mid \sigma(D_1, \dots, D_{n-1})) = 0$, where $D_i := (W_i, Y_i)$ is the data at time $i$. Although our martingale is of the form $E(f(D_n) \mid \sigma(D_1, \dots, D_{n-1})) = 0$, where $f(.)$ is the function that maps the data to $\hat\tau_{t \mid t-1} - \tau_t(w_{1:(t-1)}^{obs})$. These slight differences can complicate the original proof or Strassen's Theorem 4.4 depending on the function $f(.)$. Regardless, we can still apply Theorem 4.4 by defining a larger filtration $\tilde{\mathcal{F}}_t := \sigma(u_1, u_2, \dots, u_t)$ and apply Strassen's theorem directly to a MDS of the form $E(u_t \mid \tilde{\mathcal{F}}_{t-1})$. Note our estimand do not change even if conditioning on a larger filtration because we are in the design-based framework (hence these causal estimands are fixed regardless of what random quantity one conditions on). Regardless, this step is just a trick to apply Strassen's theorem faithfully. Specifically, we can still write this proof with respect to the original filtration $\sigma(D_1, \dots, D_{n-1})$ since $u_t$ is still a MDS with respect to this smaller filtration (once we have gotten the strong approximation we can apply the same trick as outlined below). 

The remaining conditions are identical and we omit the uniform integrability condition in Equation (138) of \citep{stratssen} since it holds trivially under our bounded potential outcome for Assumption~\ref{assumption:boundedPO_general} and our strong positivity in Assumption~\ref{assumption:boundedPO_general}.  Lastly, although this theorem uses the actual variance (not an upper bound), using an upper bound only makes the confidence sequence width strictly wider and hence the validity still holds.

Finally, utilizing Theorem 4.4 Equation (159) in \citep{stratssen}, we have that
\begin{equation}
\label{eq:strong_approx}
    \frac{1}{t} \sum_{j = 1}^t u_j = \frac{1}{t} \sum_{j = 1}^t \sigma_{j \mid j-1} Z_j + o\left(\frac{\tilde{S}_{t \mid t-1}^{3/8}\log(\tilde{S}_{t \mid t-1})}{t} \right) \quad a.s.,
\end{equation}
Combining Equation~\eqref{eq:gaussian_CS} and Equation~\eqref{eq:strong_approx} implies that with probability at least $(1 - \alpha)$, 
\begin{equation}
\label{eq:nonasymp_proofCS}
  \forall t \geq 1, \left|\frac{1}{t} \sum_{i=1}^t u_i \right| < \sqrt{ \frac{2(t\bar{\sigma}_t^2 \eta^2 + 1)}{t^2 \eta^2} \text{log}\Bigg( \frac{\sqrt{t\bar{\sigma}_t^2\eta^2 + 1}}{\alpha} \Bigg)} +  o\left(\frac{\tilde{S}_{t \mid t-1}^{3/8}\log(\tilde{S}_{t \mid t-1})}{t} \right).  
\end{equation}
Using Assumption~\ref{assumption:var_no_vanish}, we have that
\begin{equation}
\frac{1}{t} \sum_{j = 1}^t \hat\tau_{j \mid j-1} \pm \sqrt{ \frac{2(t\bar{\sigma}_t^2 \eta^2 + 1)}{t^2 \eta^2} \text{log}\Bigg( \frac{\sqrt{t\bar{\sigma}_t^2\eta^2 + 1}}{\alpha} \Bigg)}
\label{eq:true_CS}
\end{equation}
forms an $(1 - \alpha)-$asymptotic confidence sequence for $\bar{\tau}_t(w_{1:(t-1)}^{obs})$, where we used Assumption~\ref{assumption:var_no_vanish} so that the $\tilde{V}_t/V_t \xrightarrow{a.s.} 1$ holds where $\tilde{V}_t$ is the non-asymptotic confidence width in Equation~\eqref{eq:nonasymp_proofCS} (with the little $o$ term) and $V_t$ is defined in Equation~\eqref{eq:true_CS} (without the little $o$ term). 

\paragraph{Step 3: Using empirical variance} Unfortunately, the confidence sequence in Equation~\eqref{eq:true_CS} can not be directly used because $\bar{\sigma}_t$ is based off the true variance and hence not obtainable from the data. The last step is to replace Equation~\eqref{eq:true_CS} with our estimated variance $\tilde{\sigma}_t^2 \coloneqq S_{t \mid t-1}/t$. 

More precisely, we now show that if we further have $\tilde{\sigma}_t^2 \xrightarrow{a.s.} \bar{\sigma}_t^2$, then we have 
$$\frac{1}{t} \sum_{j = 1}^t \hat\tau_{j \mid j-1}  \pm \sqrt{ \frac{2(t\tilde{\sigma}_t^2 \eta^2 + 1)}{t^2 \eta^2} \text{log}\Bigg( \frac{\sqrt{t\tilde{\sigma}_t^2\eta^2 + 1}}{\alpha} \Bigg)}$$
forms a $(1-\alpha)$-asymptotic confidence sequence for $\tau_t(w_{1:(t-1)}^{obs})$, giving us the desired result. For completeness, we replicate this part of the proof under our setting. First we rewrite the assumption of $\tilde{\sigma}_t^2 \xrightarrow{a.s.} \bar{\sigma}_t^2$ as $\tilde{\sigma}_t^2 - \bar{\sigma}_t^2 = o(\bar{\sigma}_t^2)$. Then Equation~\eqref{eq:true_CS} gives us  
\begin{align*}
 \sqrt{ \frac{2(t\bar{\sigma}_t^2 \eta^2 + 1)}{t^2 \eta^2} \text{log}\Bigg( \frac{\sqrt{t\bar{\sigma}_t^2\eta^2 + 1}}{\alpha} \Bigg)} &= \sqrt{ \frac{2(t (\tilde{\sigma}_t^2 + o(\bar{\sigma}_t^2)) \eta^2 + 1)}{t^2 \eta^2} \text{log}\Bigg( \frac{\sqrt{t (\tilde{\sigma}_t^2 + o(\bar{\sigma}_t^2))\eta^2 + 1}}{\alpha} \Bigg)}  \\
 &= \sqrt{ \frac{t (\tilde{\sigma}_t^2 + o(\bar{\sigma}_t^2) ) \eta^2 + 1}{t^2 \eta^2} \text{log}\Bigg( \frac{t (\tilde{\sigma}_t^2 + o(\bar{\sigma}_t^2))\eta^2 + 1}{\alpha^2} \Bigg)} \\
 &= \sqrt{ \frac{t\tilde{\sigma}_t^2\eta^2 + o(t \bar{\sigma}_t^2)+ 1}{t^2 \eta^2} \text{log}\Bigg( \frac{t \tilde{\sigma}_t^2\eta^2 + o(t\bar{\sigma}_t^2) + 1}{\alpha^2} \Bigg)} \\
 &= \sqrt{ \left( \frac{t\tilde{\sigma}_t^2\eta^2 + 1}{t^2 \eta^2} + o(\bar{\sigma}_t^2/t) \right) \text{log}\Bigg( \frac{t \tilde{\sigma}_t^2\eta^2 + o(t\bar{\sigma}_t^2) + 1}{\alpha^2} \Bigg)}. \\
\end{align*}
Focusing on the second logarithmic term, we have
\begin{align*}
\log \Bigg( \frac{t \tilde{\sigma}_t^2\eta^2 + o(t\bar{\sigma}_t^2) + 1}{\alpha^2} \Bigg) &= \text{log}\Bigg( \frac{t \tilde{\sigma}_t^2\eta^2 + 1}{\alpha^2} + o(t\bar{\sigma}_t^2) \Bigg)\\
&= \text{log}\Bigg( \frac{t \tilde{\sigma}_t^2\eta^2 + 1}{\alpha^2} \big[1 + o(1) \big]  \Bigg)\\
&= \text{log}\Bigg( \frac{t \tilde{\sigma}_t^2\eta^2 + 1}{\alpha^2}  \Bigg) + \log(1 + o(1)) \\
&= \text{log}\Bigg( \frac{t \tilde{\sigma}_t^2\eta^2 + 1}{\alpha^2}  \Bigg) + o(1),
\end{align*}
where the last line follows because $\text{log}(1 + x) = x + o(1)$ for $|x| < 1$. Returning back to the main expression we have  
\begin{align*}
 \sqrt{ \frac{2(t\bar{\sigma}_t^2 \eta^2 + 1)}{t^2 \eta^2} \text{log}\Bigg( \frac{\sqrt{t\bar{\sigma}_t^2\eta^2 + 1}}{\alpha} \Bigg)} &= \sqrt{ \left( \frac{t\tilde{\sigma}_t^2\eta^2 + 1}{t^2 \eta^2} + o(V_t  /t^2) \right) \left[ \text{log}\Bigg( \frac{t \tilde{\sigma}_t^2\eta^2 + 1}{\alpha^2}  \Bigg)  + o(1) \right] } \\
 &= \sqrt{ \frac{t\tilde{\sigma}_t^2\eta^2 + 1}{t^2 \eta^2} \text{log}\Bigg( \frac{t \tilde{\sigma}_t^2\eta^2 + 1}{\alpha^2} \Bigg) + o(V_t/t^2) + o(V_t \log V_t/t^2 ) + o(V_t/t^2) } \\
  &= \sqrt{ \frac{t\tilde{\sigma}_t^2\eta^2 + 1}{t^2 \eta^2} \text{log}\Bigg( \frac{t \tilde{\sigma}_t^2\eta^2 + 1}{\alpha^2} \Bigg) + o(V_t \log V_t/t^2 )  } \\
 &= \sqrt{ \frac{t\tilde{\sigma}_t^2\eta^2 + 1}{t^2 \eta^2} \text{log}\Bigg( \frac{t \tilde{\sigma}_t^2\eta^2 + 1}{\alpha^2} \Bigg)} + o(\sqrt{V_t \log V_t}/t) ,
\end{align*}
where the last line follows because $\sqrt{a + b} \leq \sqrt{a} + \sqrt{b}$ for $a, b \geq 0$. This formally shows how our confidence sequence in Theorem~\ref{theorem:main_general_thm} is a valid $(1 - \alpha)-$asymptotic confidence sequence for $\mu_t$ with approximation rate $o(\sqrt{V_t \log V_t}/t)$  given that our variance estimator is strongly consistent. 

However, Theorem~\ref{lemma:moment_cond} only tells us that $\tilde{\sigma}_t^2$ is conditionally unbiased for $\bar{\sigma}_t^2$. To establish the consistency result, we again use a version of strong law of large numbers for martingale sequence difference. We denote $U_t \coloneqq  \tilde{\sigma}_t^2 - \bar{\sigma}_t^2$ and $U_t$ is a martingale sequence difference with respect to the filtration $\mathcal{F}_{t-1}$. Our goal is to show that $U_t \rightarrow 0$ almost surely. First, we define $\tilde{U}_t := \sum_{i = 1}^t U_i / i$. This quantity is bounded, a martingale-sequence difference, and also converges almost-surely because $U_t$ satisfies the desired uniform integrability by Assumption~\ref{assumption:boundedPO_general}. Finally applying Kronecker's Lemma on $\tilde{U}_t$ we get that $U_t \rightarrow 0$ almost surely. 

\section{Optimizing and choosing $\eta$ parameter}
\label{Appendix:choosing_rho}
In this section, we show in detail how an analyst can choose $\eta$ to optimize the confidence sequence width for a desired specific time $t^*$. The derivations are nearly identical to those presented in \citep{time_uniform}, but we repeat them here for completeness. 

Our confidence width presented in all the theorems have the following structure
$$B_t(\alpha) \coloneqq \sqrt{ \frac{2(t \eta^2 + 1)}{t^2\eta^2} \text{log}\Bigg( \frac{\sqrt{t \eta^2 + 1}}{\alpha}\Bigg) }, $$
where we have omitted the variance terms and instead substituted each $\hat\sigma_i = 1$ since we want $\eta$ to be data-independent.\footnote{Consequently, we are not formally optimizing $\eta$ for the actual confidence width, but $\eta$ can still be conceptually interpreted as minimizing the confidence sequence width at a desired time $t^*$ (See Appendix C.3 in \citep{time_uniform} for more details).} Then,
$$\argmin_{\eta > 0} B_t(\alpha) = \sqrt{\argmin_{x > 0} f(x)},$$
$$\text{where } \quad f(x) \coloneqq \frac{t  x + 1}{t^2 x} \text{log}\left(\frac{t x + 1}{\alpha^2} \right), \quad x \coloneqq \eta^2.$$
Furthermore, $\lim_{x \rightarrow 0} f(x) = \lim_{x \rightarrow \infty} f(x)$ and thus if we can find the critical point by finding a solution for $\partial{f}/\partial{x} = 0$, then this must be the unique minimum.

Therefore, we have that 
$$\frac{\partial{f}}{\partial{x}} = -\frac{1}{t^2 x^2}\text{log}\left(\frac{tx + 1}{\alpha^2} \right) + \frac{1}{tx}.$$
Setting the above to zero, we obtain
$$-\alpha^2 \text{exp}(1) = -(tx + 1) \text{exp}(-(tx + 1)).$$
Therefore, we have that the solution is $-(tx + 1) = W_{-1}(-\alpha^2 \text{exp}(1))$, where $W_{-1}$ is the lower branch of the Lambert $W$ function. The solution only exists if 
$$-\alpha^2 \text{exp}(1) \geq -\text{exp}(1),$$
or equivalently if $\alpha^2 \leq 1$, which is always true for any $\alpha \in [0, 1]$. Therefore, we have that 
$$\argmin_{\eta > 0} B_t(\alpha) = \sqrt{\frac{-W_{-1}(-\alpha^2 \text{exp}(1)) - 1}{t^* }}.$$

\newpage
\section*{Appendix References}
\putbib 
\end{bibunit}

\end{document}